\documentclass[aps,amsmath,amssymb,onecolumn,superscriptaddress]{revtex4-1}

\usepackage{url}
\usepackage{mathtools}
\usepackage{graphicx}
\usepackage{xcolor}
\usepackage{subcaption}

\newcommand{\dint}[1]{\mathrm{d}{#1}}
\newcommand{\ddif}[2]{\frac{\mathrm{d}{#1}}{\mathrm{d}{#2}}}
\newcommand{\dpart}[2]{\frac{\mathrm{\partial}{#1}}{\mathrm{\partial}{#2}}}
\newcommand{\dparth}[3]{\frac{\mathrm{\partial}^{#3}{#1}}{\mathrm{\partial}{#2}^{#3}}}

\newcommand{\fe}{\mathcal{F}}

\newcommand{\kBT}{k_\mathrm{B} T}	
\newcommand{\kbT}{k_\mathrm{B} T}

\newcommand{\Pe}{\mathrm{Pe}}

\newcommand{\Ca}{\mathrm{Ca}}

\date{\today}

\begin{document}
\title{Periodic phase-separation during meniscus-guided deposition}
	\author{René de Bruijn}
	\email{r.a.j.d.bruijn@tue.nl}
	\affiliation{Department of Applied Physics and Science Education, Eindhoven University of Technology, P.O. Box 513, 5600 MB Eindhoven, The Netherlands}
	\affiliation{Institute for Complex Molecular Systems, Eindhoven University of Technology, P.O. Box 513, 5600 MB Eindhoven, The Netherlands}
 	\author{Anton A. Darhuber}
	\affiliation{Department of Applied Physics and Science Education, Eindhoven University of Technology, P.O. Box 513, 5600 MB Eindhoven, The Netherlands}
 	\author{Jasper J. Michels}
	\affiliation{Max Planck Institute for Polymer Research, Mainz, Germany}
  	\author{Paul van der Schoot}
	\affiliation{Department of Applied Physics and Science Education, Eindhoven University of Technology, P.O. Box 513, 5600 MB Eindhoven, The Netherlands}
 
\begin{abstract}
We numerically investigate the meniscus-guided coating of a binary fluid mixture containing a solute and a volatile solvent that phase separates via spinodal decomposition. Motivation is the evaporation-driven deposition of material during the fabrication of organic thin film electronics. We find a transition in the phase-separation morphology from an array of droplet-shaped domains deposited periodically parallel to the slot opening to isotropically dispersed solute-rich droplets with increasing coating velocity. This transition originates from the competition between the hydrodynamic injection of the solution into the film and diffusive transport that cannot keep up with replenishing the depletion of solute near the solute-rich domains. The critical velocity separating the two regimes and the characteristic length scale of the phase-separated morphologies are determined by the ratio of two emergent length scales: (i) the spinodal length, which implicitly depends on the evaporation rate and the properties of the solution, and (ii) a depletion length proportional to the ratio of the tracer diffusivity of the solute and the coating velocity. For coating below the critical velocity, an array of droplet-shaped domains is deposited periodically parallel to the slot opening, with the domain size and deposition wavelength proportional to a solute depletion length. As the competition in the mass transport is inherent in any kind of unidirectional deposition of demixing solutions, our findings should apply to a broad range of coating techniques and forced demixing processes.
\end{abstract}
\maketitle

\section{Introduction}
Over the past few decades, thin film organic electronics have emerged as a promising alternative to silicon-based electronics owing to their ease-of-manufacturing and favorable electrical, mechanical and optical properties~\cite{Katz2009Thin-FilmDevices,Yin2020HighlyDisplays}. Typical applications range from batteries and displays to transistors and organic photovoltaic devices~\cite{Mei2013,Ling2018,Janssen2019}. The active thin film typically consists of (a blend of) polymeric or small organic molecules~\cite{Mei2013,Janssen2019}. This film is manufactured from solution containing the active components dissolved in one or more volatile solvents, deposited on a substrate and dried via the evaporation of the solvent~\cite{Richter2017MorphologyViewpoint}. The final dry film has a complex microscopic morphology that emerges spontaneously, for instance via liquid-liquid phase separation, crystallization of one or more of the components, or a combination of both~\cite{Peng2023ASemiconductors}. Obtaining the correct morphology is essential for the efficient functioning of the devices~\cite{Gaspar2018RecentCells,McDowell2018SolventCells}. 

\begin{figure}[htb]
    \centering
    \includegraphics[width=\columnwidth]{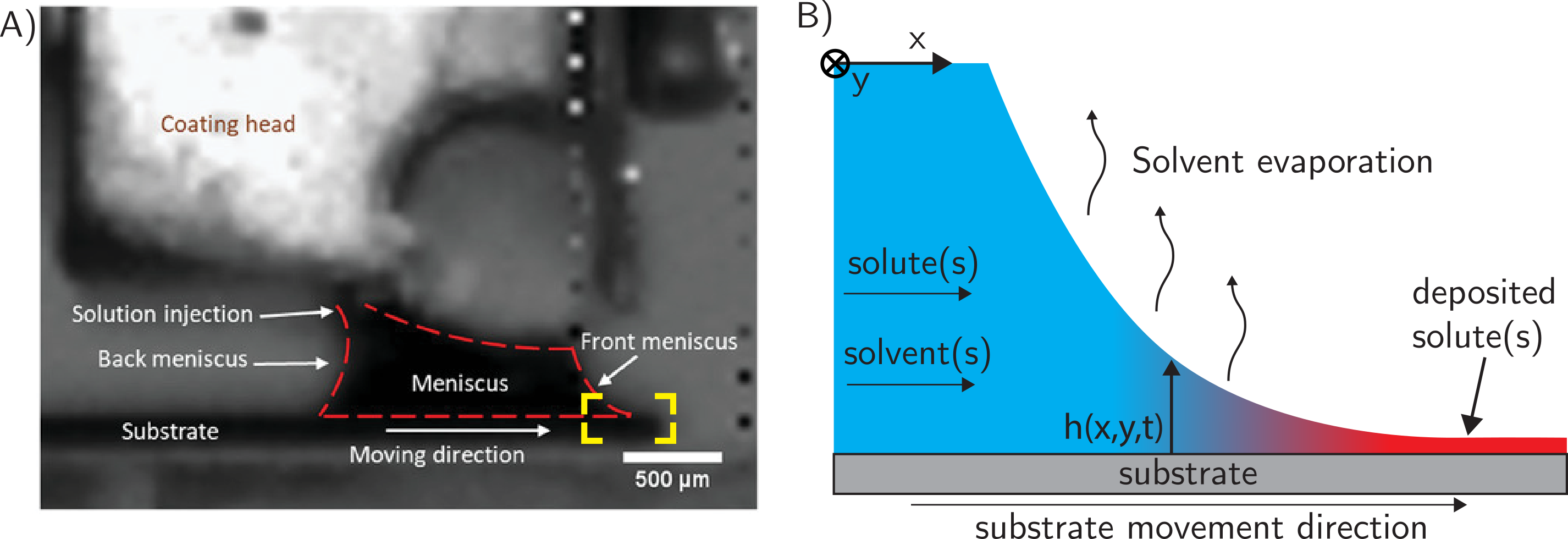}
    \caption{A) A side-view image of a zone-casting setup reproduced from Ref.~\cite{Yildiz2022OptimizedCoating}. The coating head deposits the solution on the substrate that moves from left to right, resulting in a liquid meniscus highlighted by the red dashed line between the coating head and the substrate. A wet or dry thin film entrains from the front meniscus, not observable on the scale of the figure. B) A schematic view of the region between the yellow dashed box in A. The solution inlet is on the left, and solvent evaporates out of the solution and dries the thin film. The colors indicate the local volume fraction of the solute, with blue indicating low solute concentration and red high solute concentration. Our coordinate system is depicted in the upper left corner for future reference and the height $h$ of the solution-gas interface is indicated. The figure is not to scale.}
    \label{fig:experimental_setup}
\end{figure}

Meniscus-guided coating (MGC) represents a family of deposition techniques, where the solution is deposited from a stationary dispensing unit onto a moving substrate. A typical zone-casting set up~\cite{Yildiz2022OptimizedCoating} is illustrated in Fig.~\ref{fig:experimental_setup}. In contrast to spin-coating, MGC techniques are scalable and enable large-area manufacturing. Zone casting in particular allows for optimal control over the deposition of the solution and hence the emerging thin-film morphology~\cite{Gu2018ThePolymers}. For the particular setup shown in Fig.~\ref{fig:experimental_setup}A, a (pre-heated) solution is injected from a coating head onto the moving substrate, resulting in a liquid meniscus (red dashed curve) between the substrate and the coating head. The thin film entrains from the front meniscus due to the direction of motion of the substrate and dries via solvent evaporation, as shown schematically in Fig.~\ref{fig:experimental_setup}B. The front meniscus can be divided into two regions (not shown in Fig.~\ref{fig:experimental_setup}B: the so-called static meniscus, which is the region of space wherein the meniscus shape is independent of the substrate velocity and further removed from the coating head is the dynamic meniscus, the shape of which, as the name suggests, does depend on the coating velocity. The properties of the film, such as its thickness, the structure of the free surface and the compositional morphology can be controlled via the fine-tuning of the material properties of the mixture and the processing conditions~
\cite{Cao2018HowPhotovoltaics,Levitsky2020TowardBlends}. Experimentally, the rate of solvent evaporation~\cite{Franeker2015spincoating,Negi2018SimulatingInvestigation,Schaefer2016StructuringEvaporation}, the substrate velocity and the meniscus shape~\cite{Zhang2020KeyCoating,Yildiz2022OptimizedCoating} have all been identified as important parameters controlling the dry film properties. 

\begin{figure}[tb]
    \centering
    \includegraphics[width=236.84843pt]{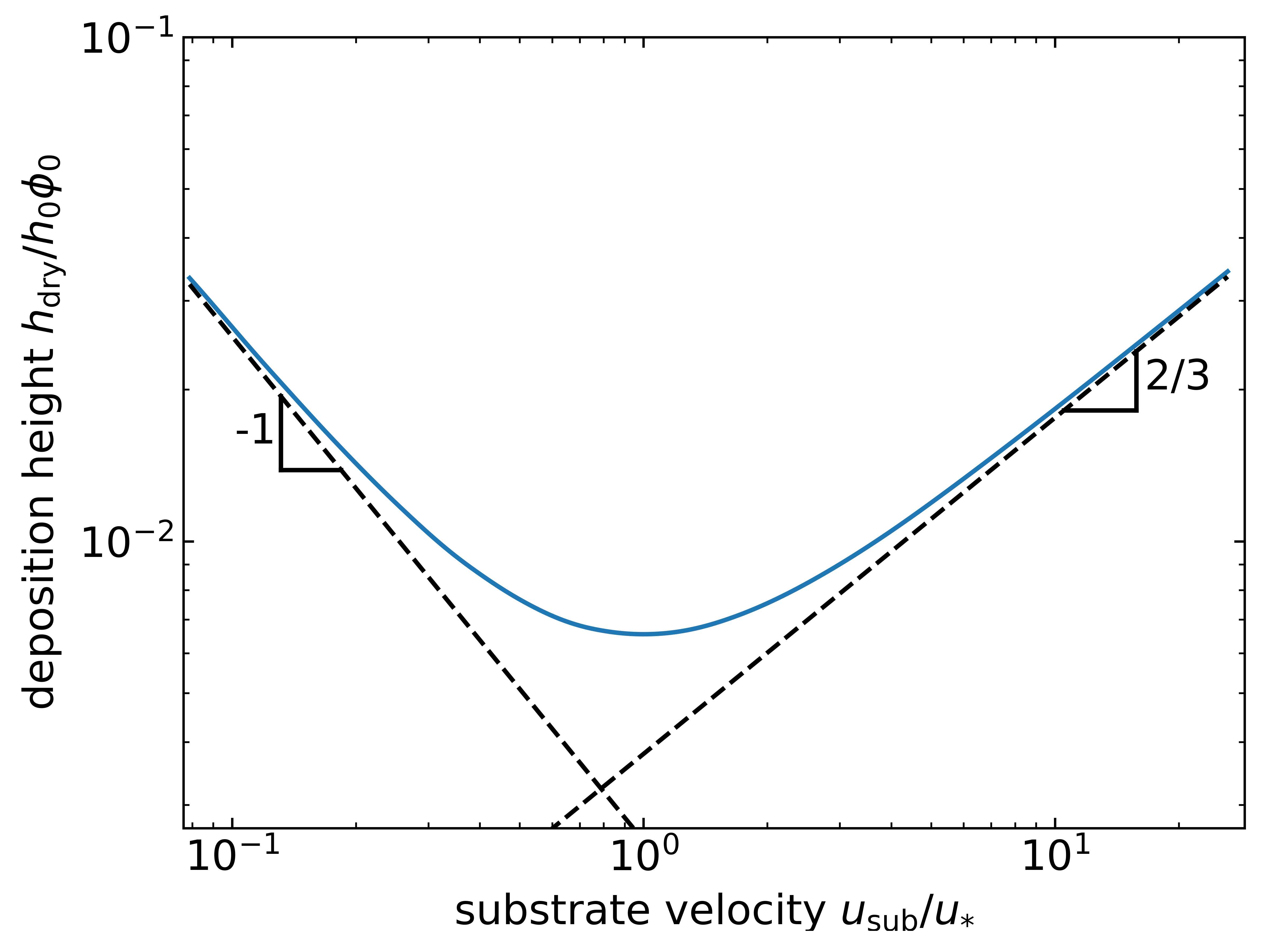}
    \caption{The deposited dry film height $h_\mathrm{dry}$ divided by the height at the inlet $h_0$ and the volume fraction at the inlet $\phi_0$ for a regular solution with Flory interaction parameter $\chi = 0$, as a function of the substrate velocity $u_\mathrm{sub}/u^{}_{*}$ divided by the critical velocity $u^{}_{*}=150$ \textmu m s$^{-1}$ for which the height attains its minimum value. For the physical parameters we use the values given in Table~\ref{tab:ParameterRanges}. The dashed lines represent the expected scaling exponents for low substrate velocities $h_\mathrm{dry} \sim u_\mathrm{sub}^{-1}$ in the evaporative regime and high substrate velocities $h_\mathrm{dry} \sim u_\mathrm{sub}^{2/3}$ in the Landau-Levich regime.}
    \label{fig:Evap_LandauLevich}
\end{figure}

What controls the thickness or height of the dry film is reasonably well-understood~\cite{LeBerre2009FromThickness}: two coating regimes are known to emerge as a function of the coating velocity~\cite{LeBerre2009FromThickness,Doumenc2010DryingPhases}, illustrated in Fig.~\ref{fig:Evap_LandauLevich}. In the so-called evaporative regime, where evaporation is in some sense fast relative to the coating process, a dry film entrains from the meniscus and mass conservation dictates that the height $h_\mathrm{dry}$ is inversely proportional to the coating velocity $u_\mathrm{sub}$, $h_\mathrm{dry}\sim u_\mathrm{sub}^{-1}$~\cite{Jing2010DryingNumber}. If on the other hand evaporation is comparatively slow, in the so-called Landau-Levich regime, a wet film is deposited that subsequently dries. The dry film height is then set by a balance of viscous and capillary forces, in which case $h_\mathrm{dry} \sim u_\mathrm{sub}^{2/3}$~\cite{Landau1988DraggingPlate}. At the critical velocity $u^{}_{*}$ that separates these two regimes the dry film height attains its minimum value. Note that while the two scaling regimes are universal, the prefactors differ and therefore the dry height curve is not universal. The structural properties of the free surface of the film are also reasonably well-understood in terms of dewetting at the contact line and Marangoni-instabilities~\cite{Doumenc2010DryingPhases,Frastia2012ModellingSuspensions,Doumenc2013Self-patterningMeniscus}. This has been studied extensively in experiments, simulations and theory, showing that (periodic) patterns for the free film surface form even if the composition is homogeneous or varies only over large distances~\cite{Yabu2005PreparationMeniscus,Frastia2012ModellingSuspensions,Chen2012QuasiSolution,Wedershoven2018PolymerConvection,Doumenc2013Self-patterningMeniscus}, ranging from (nearly) flat films to striped surface patterns both parallel and perpendicular to the coating direction with wavelengths typically on the order of 100 - 1000 \textmu m~\cite{Dey2016NumericalRegime}, and periodically deposited droplet-like domains~\cite{Yabu2005PreparationMeniscus,Ly2020Two-dimensionalPath,Doumenc2013Self-patterningMeniscus,Dey2016NumericalRegime,Wedershoven2018PolymerConvection}.

In contrast to the height of the deposited film, what controls the emerging structure in the \textit{composition} of the film in is poorly understood, in particular in case of phase-separating solutions. The existing experimental studies clearly demonstrate that the meniscus shape and the coating velocity strongly influence the resulting compositional morphology in the dry film~\cite{He2017Meniscus-assistedCells,Janneck2019InfluenceFilms,Chen2020UnderstandingFabrications,Zhang2020KeyCoating,Zhang2021RelationFilms,Wu2022RecentApproaches,Yildiz2022OptimizedCoating}, yet how exactly remains unclear in large part due to the non-trivial coupling between the process settings and the dynamics of the involved phase transition(s). Improved model-based understanding of the correlation between the process parameters and the final dry film morphology is essential for optimizing coating conditions and enhancing morphological control. Recently, Michels \textit{et al.}~\cite{Zhang2021RelationFilms,Michels2021PredictiveCoating} studied numerically the meniscus-guided deposition of a crystallizing solute from a volatile solution and found qualitative correspondence between calculated and experimental film morphologies~\cite{Zhang2021RelationFilms} in both the evaporative and the Landau-Levich regimes. However, since they neglect hydrodynamic transport, they modeled the deposition behavior in the aforementioned coating regimes by artificially imposing the expected thickness scaling using an empirical parametrization based on experimental data. What is more, they modeled the translation of the meniscus implicitly via their solvent evaporation model~\cite{Michels2021PredictiveCoating}. In general, this simplification is justifiable when coating is relatively fast because the drying of the film then decouples from the coating process. However, it becomes inaccurate in the low-velocity evaporative regime where the two processes are coupled~\cite{Doumenc2010DryingPhases}. It is especially at these low velocities, typically below a hundred micrometers per second~\cite{Zhang2020KeyCoating}, that the emerging structure correlates strongly with the coating velocity and other hydrodynamic transport processes~\cite{Zhang2021RelationFilms,Michels2021PredictiveCoating}. Hence, accounting for the hydrodynamic transport in the meniscus and the film is a necessary ingredient of a fully predictive theory of meniscus-guided morphological evolution in solutions.

In this work, we study the correlation between the control parameters relevant to meniscus-guided deposition and the emergent structure in the composition of a binary solution that liquid-liquid phase separates as it is being coated. We limit our study to the early stages of demixing and do not study in detail the late-time ripening of the morphology, even though it does occur in the model. Two distinct and competing phase-separation length scales emerge in the coating process: (i) a spinodal length scale that somehow results from a quench due to solvent evaporation~\cite{Schaefer2016StructuringEvaporation,Negi2018SimulatingInvestigation} and (ii) a depletion length scale set by the competition between hydrodynamic transport of solute molecules from the meniscus into the film and diffusive depletion of solute near the phase-separated domains. The ratio of these two length scales turns out to determine a critical velocity. For deposition velocities slower than this critical velocity, the deposition structure changes into a periodic deposition of an array of elongated solute-rich domains aligned with the direction of substrate movement. For faster deposition, the structure eventually becomes similar to that of spinodal decomposition in a flat, stationary film, wherein the droplets are dispersed isotropically. 

The remainder of this paper is set up as follows. In Section~\ref{sec:theory} we introduce our model and discuss the methodology for our numerical calculations in Section~\ref{sec:numerics}. In Section~\ref{sec:twodim}, we study the deposition pattern using vertically-averaged numerical calculations, showing elongated domain growth for low substrate velocities and a (nearly) deterministic periodic deposition of an array of droplet-shaped domains whose centers-of-masses lie on a line oriented perpendicular to the coating direction. We study the relevant length scales in more detail using quasi one-dimensional calculations in Section~\ref{sec:onedim}, in which we ignore any structural patterning in the directions perpendicular to the deposition direction. Section~\ref{sec:initialtransient} focuses on the effect of the initial conditions in our numerical calculations. In Section~\ref{sec:discussion} we discuss and summarize our work.

\section{Model}\label{sec:theory}
\begin{figure}[tb]
    \centering
    \includegraphics[width=0.49\columnwidth]{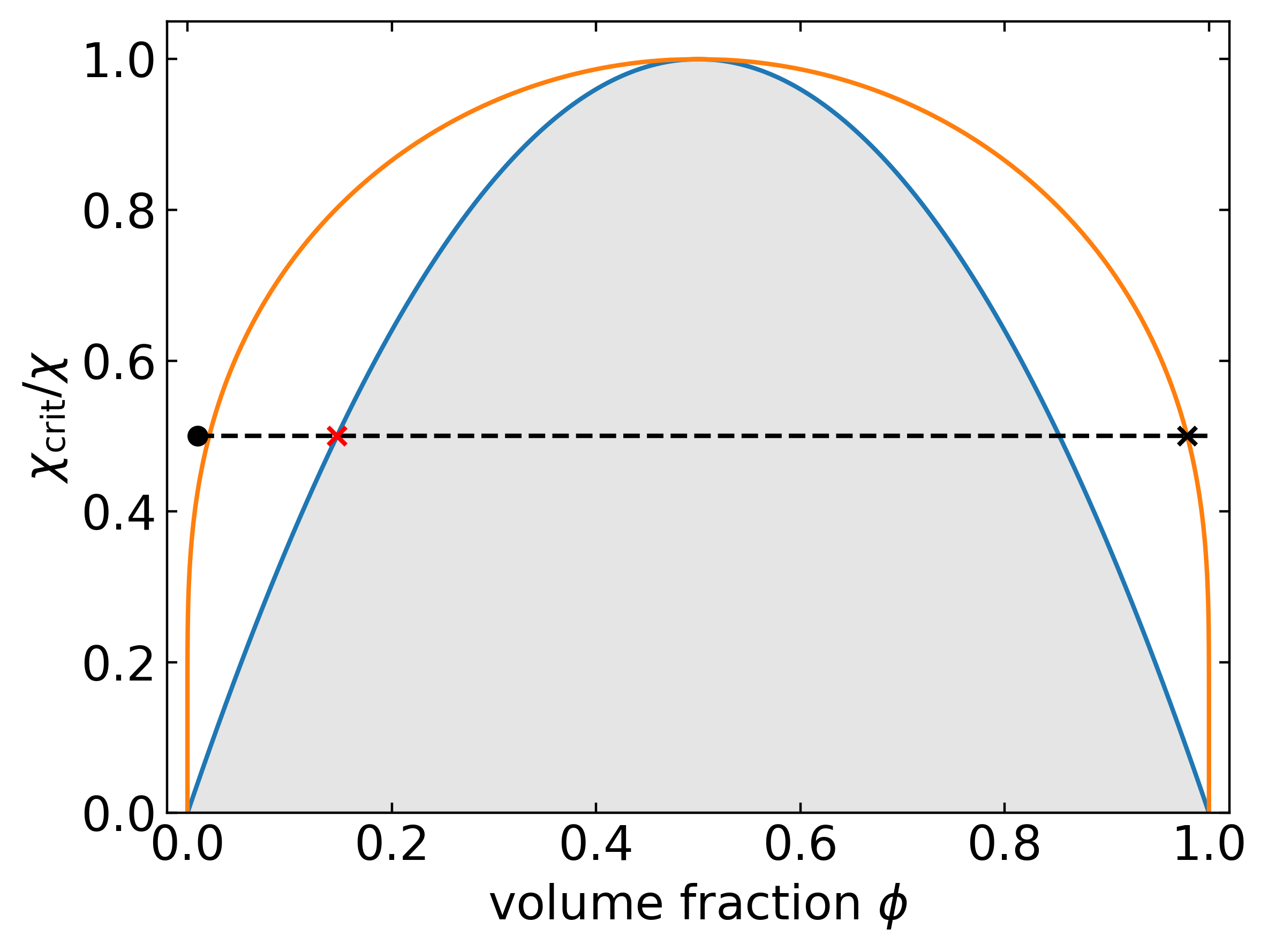}
    \caption{The phase diagram of a binary solution within the regular solution model. We plot the inverse interaction strength $\chi_\mathrm{crit}/\chi$ with $\chi_\mathrm{crit} =2$ as a function of the volume fraction. The orange line indicates the binodal line or metastability limit and the blue line the spinodal line. For volume fractions between the binodal and spinodal lines, the solution is metastable and can phase separate via the nucleation-and-growth mechanism. For any state between the spinodals, shaded in gray, the solution is unstable and demixes via spinodal decomposition. The solution is stable otherwise. The dashed line is the path traversed (from left to right) in our calculations due to solvent evaporation. The black circle is the initial concentration, the red cross indicates the phase state where the spinodal line is crossed, the black cross indicates where the phase state where the binodal is traversed and the phase-separated domains redissolve. We refer to the collection of states between crossing the low-concentration spinodal and the high-concentration binodal as the immiscible region.}
    \label{fig:Phase_Diagram}
\end{figure}

Let us consider the isothermal meniscus-guided deposition of an incompressible binary solution consisting of a solute and a volatile solvent onto a moving substrate, as shown schematically in Fig.~\ref{fig:experimental_setup}B, mimicking, \textit{e.g.}, zone casting~\cite{Tang2011RobustCasting,Zhang2020KeyCoating}. The solution is deposited while still in a homogeneous state but is subsequently quenched beyond the spinodal concentration due to solvent evaporation~\cite{Kouijzer2013PredictingBlends}, resulting in the formation of regions rich and poor in solute near the drying front~\cite{Franeker2015cosolvents,Franeker2015spincoating}. We schematically indicate this evaporation-driven demixing using the phase diagram depicted in Fig.~\ref{fig:Phase_Diagram} as a function of the inverse Flory interaction parameter with $\chi_\mathrm{crit} = 2$ and the solute volume fraction $\phi$. The binodal and spinodal lines are depicted with the orange and blue curves, respectively. The initial state is indicated with the black dot, and the path (from left to right) traversed through the phase diagram due to solvent evaporation is indicated with the dashed black line. Due to the slowness of nucleation, the metastable region between the binodal and spinodal lines is typically traversed and phase separation commences only after crossing into the unstable region at the red cross~\cite{Franeker2015cosolvents,Franeker2015spincoating}. The unstable region is shaded in gray. The solution redissolves at the black cross. Redissolution is, of course, an artifact of the binary solution, whereas typical mixtures contain multiple solutes which solidify in a phase-separated state~\cite{Franeker2015spincoating}.

The standard approach to describe the (phase-separation) dynamics in fluids is based on the (bulk) Navier-Stokes-Korteweg equations in conjunction with a generalized diffusion-advection equation, also known as model H~\cite{Hohenberg1977TheoryPhenomena}. In this work, we treat both within the lubrication approximation~\cite{Oron1997Long-scaleFilms,Thiele2016GradientSurfactant}, averaging both sets of equations over the height of the film. This reduces the dimensionality of the problem from three to quasi two dimensions, because the height of the film is still modeled explicitly. This is allowed if (i) the slope of the height $h$ of the film is small, implying that $|\nabla h| \ll 1$, and (ii) (collective) diffusion is sufficiently fast, so that appreciable stratification in the solution layer does not occur~\cite{Larsson2022QuantitativeFilms}. Hydrodynamic transport is described by the lubrication equation~\cite{Oron1997Long-scaleFilms,Thiele2016GradientSurfactant,Naraigh2010NonlinearFilms}
\begin{equation}\label{eq:Lubrication}
\dpart{h}{t} + \nabla\cdot \left(h \mathbf{u}\right) = -g(\phi).
\end{equation}
Here, $h\equiv h(x,y,t)$ is the height of film in the $x$--$y$ plane of the solid substrate, or, in other words, the distance between the solid substrate and the interface between the solution and the gas phase, $\nabla = (\partial_x,\partial_y)$ is the in-plane gradient vector and $ \mathbf{u} =  \mathbf{u}(x,y,t) $ the height-averaged fluid velocity vector. The function $g$ accounts for the evaporation of the solvent and is a function of the height-averaged local volume fraction $\phi = \phi(x,y,t)$ of the solute, specified below. We neglect inertia in Eq.~\eqref{eq:Lubrication} as Reynolds numbers for thin film flows are generally much smaller than unity~\cite{Oron1997Long-scaleFilms}. 

We write the height-averaged fluid velocity in Eq.~\eqref{eq:Lubrication} as a sum of three terms,
\begin{equation}\label{eq:HavgVelo}
\mathbf{u} = \mathbf{u}_\mathrm{sub} - \frac{h^2}{3 \eta}\nabla p  - \frac{\kBT \kappa h}{3 \eta b^{3}} \nabla \left(h |\nabla \phi|^2\right).
\end{equation}
The first term in Eq.~\eqref{eq:HavgVelo} represents the velocity of the substrate, assuming no-slip conditions. In our coordinate system, see Fig.~\ref{fig:experimental_setup}B, we define the $x$-direction to be parallel to the coating direction, so $\mathbf{u}_\mathrm{sub} = (u_\mathrm{sub},0)^{T}$ with $u_\mathrm{sub} = |\mathbf{u}_\mathrm{sub}|$. The other two contributions represent the response to the forces that act on the film. The second term accounts for gradients in the pressure $p$~\cite{Oron1997Long-scaleFilms}, where $\eta$ denotes the viscosity of the solution, which, in principle, should depend on the local concentration of solute. We choose to ignore this for reasons of simplicity. The pressure itself has in our model two contributions, 
\begin{equation}\label{eq:pressure}
    p = -\sigma\nabla^2 h - \frac{A_\mathrm{H}}{6 \pi h^3}.
\end{equation}
The first contribution in Eq.~\eqref{eq:pressure} describes the capillary pressure originating from the curvature of the liquid-gas interface and the second the impact of the so-called disjoining pressure~\cite{LIFSHITZ1992TheSolids,Dai2008DisjoiningFilms}. The capillary pressure is a function of the interfacial tension $\sigma$ of the gas-liquid interface that we presume composition-independent. In our isothermal model we neglect thermal or compositional Marangoni stresses that have their origin in gradients in the interfacial tension with the gas phase. 
The disjoining pressure is proportional to the Hamaker constant $A_\mathrm{H}$ describing the strength of the Van der Waals interaction between the solid and the fluid relative to that between the gas and the solid. 
We neglect any corrections to the disjoining pressure stemming from gradients in the height of the film~\cite{Dai2008DisjoiningFilms}. 
The third and final contribution to Eq.~\eqref{eq:HavgVelo} accounts for the capillary pressure originating from the curved interfaces between different (liquid) phases in the solution itself. Within the diffuse interface description, these are commonly referred to as Korteweg stresses~\cite{Anderson1998DIFFUSE-INTERFACEMECHANICS,Kim2012Phase-FieldFlows,Jasnow1996Coarse-grainedFlow,Hohenberg1977TheoryPhenomena}. In this term  $\kBT$ denotes the thermal energy with $k_\mathrm{B}$ Boltzmann's constant and $T$ the absolute temperature, $b^{3}$ is a microscopic volume that enters our problem for dimensional consistency and $\kappa$ the so-called ``stiffness'' of a concentration gradient, the interpretation of which becomes apparent below. 

Solvent evaporation we account for in Eq.~\eqref{eq:Lubrication} using
\begin{equation}\label{eq:evapmodel}
    g(\phi) = 
    \begin{cases}
    k \left(\phi_\mathrm{max} - \phi\right) &\text{if $\phi < \phi_\mathrm{max}$,} \\
    0 &\text{if $\phi \geq \phi_\mathrm{max}$.}\\
    \end{cases}
\end{equation}
Here, $k$ is a mass transfer coefficient, $\phi$ is as before the solute volume fraction and $\phi_\mathrm{max}\leq 1$ a cut-off for solvent evaporation. This cut-off value ensures that the (dry) film is not a pure solute phase, which, within our model, turns out to result into a diverging driving force for diffusion which makes our numerical approach slow down excessively. We set the cut-off $\phi_\mathrm{max}$ to a value higher than the high volume fraction branch of the binodal as it may otherwise affect the phase-separation kinetics. For our purpose, setting $\phi_\mathrm{max} = 0.999$ suffices to suppress any numerical issues.

Next, we focus on the mass transport of the solute in the thin film, which we model using the vertically-averaged diffusion-advection equation. This is reasonable if (i) diffusion is sufficiently fast to suppress solute accumulation at the solution-air interface due to solvent evaporation as already alluded to~\cite{Larsson2022QuantitativeFilms}, (ii) preferential interactions of the surfaces with one of the components in the solution are sufficiently weak~\cite{Naraigh2010NonlinearFilms} and (iii) the characteristic (phase-separated) domain size is larger than the height of the film. Assumption (i) is warranted if $k h/D_\mathrm{coop} < 1$, with $D_\mathrm{coop}$ the cooperative diffusion coefficient~\cite{Schaefer2017DynamicSolutions}. We ensure that this condition holds in our calculations. We neglect any preferential substrate interactions in this work, which could otherwise result in stratified or ``surface-directed'' phase separation~\cite{Puri1994Surface-directedSimulation,Naraigh2010NonlinearFilms}. We shall justify assumption (iii) \textit{a posteriori}.

The diffusion-advection equation describing the redistribution of solute in the film reads within the lubrication approximation~\cite{Thiele2016GradientSurfactant}
\begin{equation}\label{eq:genDifEq}
\dpart{\psi}{t} + \nabla \cdot \left(\psi\mathbf{u} \right) = -\nabla \cdot \mathbf{J}_\mathrm{diff} +  \zeta.
\end{equation}
Here, $\psi = \psi(x,y,t) = \phi(x,y,t) h(x,y,t)$ denotes the equivalent height of the dry solute in the solution in unit of meter, henceforth referred to as the ``solute height''~\footnote{Perhaps counter-intuitively, the solute height $\psi$, and not the volume fraction $\phi$, is the relevant independent variable. This becomes evident if we consider a hypothetical solution ``column'' of height $h$ and volume fraction $\phi$. If we (somehow) vary (in the variational sense) the height $h$ for a fixed value of the volume fraction $\phi$, or equivalently, vary the volume of the column, we must allow for the number of solute or solvent molecules in the column to change to conserve the volume fraction. Hence, $\phi$ and $h$ cannot be independent variables.}, and $\mathbf{u} = \mathbf{u}(x,y,t)$ is again the height-averaged velocity defined in Eq.~\eqref{eq:HavgVelo}. The right-hand-side of Eq.~\eqref{eq:genDifEq} accounts for diffusive transport with $\mathbf{J}_\mathrm{diff}$ the collective diffusive mass flux and $\zeta=\zeta(x,y,t)$ thermal noise. 

Using Onsager's reciprocal relations, the bare diffusive mass fluxes for both the solute and solvent component can be related to gradients in the chemical potential $\mathbf{J}_{\mathrm{diff},i}= -\sum_{j} \Lambda_{ij} \nabla \mu_j$, with $\Lambda_{ij}$ Onsager's mobility coefficients and $i,j \in [1,2]$ numerical labels for the solute and solvent components~\cite{Onsager1931ReciprocalI.,Onsager1931ReciprocalII.}. Note that the chemical potential is actually a chemical potential \textit{density} (in units of Joule per cubic meter), because of how we define it below. Following the standard approach, we relate the mobility coefficients to the tracer diffusion coefficients by demanding that the diffusion is Fickian in the dilute limit and neglect cross-mobilities ($\Lambda_{ij} =0$ for $i\neq j$)~\cite{deGennes1980DynamicsBlends,Binder1983CollectiveMixtures}. Only one of the mobility coefficients is an independent parameter due to the imposed incompressibility of the solution, implying we can express the collective diffusive flux in Eq.~\eqref{eq:genDifEq} in terms of the difference in the chemical potential density between the solute and solvent, known as the exchange chemical potential density $\Delta \mu$~\cite{Schaefer2016StructuringEvaporation}. The collective diffusive flux is then given by 
\begin{equation}
\mathbf{J}_\mathrm{diff} = -h M(\phi)\nabla\Delta \mu,
\end{equation}
with $M(\phi)$ a mobility describing interdiffusion~\cite{Thiele2016GradientSurfactant,deGennes1980DynamicsBlends,Binder1983CollectiveMixtures}. We first discuss the mobility $M$ and subsequently the exchange chemical potential density $\Delta \mu$.

The mobility $M$ is related to tracer diffusivity of both the solute and solvent, since the incompressibility of the solution means that molecules must interchange position in order to move (so-called ``interdiffusion''). The resulting expression for the mobility $M$ depends on how precisely incompressibility is enforced. The limiting cases correspond to either what are known as the ``slow-mode'' and the ``fast-mode'', referring to whether the slowest or the fastest component dictates interdiffusion~\cite{deGennes1980DynamicsBlends,Binder1983CollectiveMixtures}. Here, we consider only the limit where both components have an identical tracer diffusivity and are of similar molecular size, in which case the mobility $M$ follows trivially from Onsager's reciprocal relations~\cite{Onsager1931ReciprocalI.,Onsager1931ReciprocalII.,deGennes1980DynamicsBlends} 
\begin{equation}\label{eq:doubledeg}
    M = \frac{b^{3}}{\kBT} D \phi(1-\phi),
\end{equation}
where $D$ is the tracer diffusivity that we assume to be constant, and as before $\kbT$ the thermal energy and $b^{3}$ a microscopic volume that enters our problem for dimensional consistency.

The exchange chemical potential \textit{density} provides the thermodynamic driving force for diffusive mass transport and derives from a thermodynamic description of the excess free energy functional $\Delta\fe$ in terms of a functional derivative with respect to the solute height $\psi$ as $\Delta \mu = \delta \Delta\fe/\delta \psi$~\footnote{Incompressibility is incorporated in our expression for the free energy excess. Hence, the derivative with respect to the solute height, by construction, results in the exchange chemical potential density.}. Here, we take the variations with respect to the solute height $\psi$ instead of the volume fraction $\phi$ because the latter implicitly depends on the height of the film $h$ via $\phi \sim h^{-1}$ and therefore $\Delta \mu$ should be defined via either a constrained variation with respect to $\phi$~\cite{Clarke2005TowardMixtures}, or equivalently, via a variation with respect to $\psi$~\cite{Thiele2016GradientSurfactant}. As alluded to above, we presume that the solution remains vertically uniform and invoke the quasi two-dimensional square-gradient free energy functional for a solution of height $h$~\cite{Cahn1958FreeEnergy,Cahn1959FreeFluid,deGennes1980DynamicsBlends,Clarke2005TowardMixtures,Thiele2016GradientSurfactant},
\begin{equation}\label{eq:FreeEnergy}
\frac{\Delta\fe[h,\phi]}{\kBT} =b^{-3} \int_{-\infty}^{\infty}\dint y\int_{0}^{\infty}\dint x h \left(f_\mathrm{loc}(\phi) + \frac{\kappa}{2} |\nabla \phi|^2 + f_\mathrm{num}(\phi)\right).
\end{equation}
Here, the integration is implied over the half-plane area of the film over positions downstream of the inlet at $x = 0$, $f_\mathrm{loc}$ is the dimensionless local free energy density, $\kappa$ the earlier-introduced gradient stiffness and $f_\mathrm{num}(\phi)$ a phenomenological term that penalizes the emergence of very pure phases, the reason for which becomes apparent below.

For $f_\mathrm{loc}$ we use the Flory-Huggins free energy density for a binary solution
\begin{equation}\label{eq:locFreeEnergyDenst}
f_\mathrm{loc}(\phi) =\phi\ln \phi+ \left(1-\phi\right)\ln (1-\phi) + \chi \phi (1-\phi),
\end{equation}
where $\chi$ is the Flory-Huggins interaction parameter~\cite{Huggins1942SomeCompounds.,Flory1942ThermodynamicsSolutions,Flory1981PrinciplesPhysics}. Extending this description to polymeric solutes is straightforward and we return to this at the end of this manuscript. Since we treat the solute as a low molecular weight compound, we disregard the molecular size disparity that would otherwise enter Eq.~\eqref{eq:locFreeEnergyDenst} and we treat the stiffness $\kappa$ as a constant even though it is proportional to the Flory-Huggins interaction parameter $\chi$, as in fact is customary in the phase-field community~\cite{Schaefer2016StructuringEvaporation,Negi2018SimulatingInvestigation,Konig2021Two-dimensionalMixtures,Ronsin2022PhaseFieldFilms}. For polymeric solutions, $\kappa$ also becomes a function of the local concentration $\phi$ and the molecular weight~\cite{Debye1959AngularMixtures,McMaster1975AspectsSystems,deGennes1980DynamicsBlends}. Based on these considerations, we find the exchange chemical potential density to read $\Delta \mu = \left(\kBT/b^{3}\right) \left(\partial_\phi f_\mathrm{loc}(\phi) + \partial_\phi f_\mathrm{num}(\phi) - h^{-1} \kappa \nabla \cdot h \nabla \phi\right)$. For (very) pure phases the exchange chemical potential density based on Eq.~\eqref{eq:locFreeEnergyDenst} diverges, which turns out to slow down our numerical time integrator. The phenomenological term $f_\mathrm{num}(\phi)$ in Eq.~\eqref{eq:FreeEnergy} adds a free energetic penalty for volume fractions close to zero or unity, effectively suppressing the appearance of domains of very pure phases. This term is standard in the field and does not appreciably affect the thermodynamics and kinetics of phase-separation~\cite{Negi2018SimulatingInvestigation,Ronsin2022PhaseFieldFilms}. See Appendix~\ref{app:numerics} for a discussion.

The final ingredient in Eq.~\eqref{eq:genDifEq} is the Gaussian thermal noise term $\zeta$. As usual, we presume equilibrium moments $\langle\zeta(\mathbf{r},t)\rangle = 0$ and $\langle \zeta(\mathbf{r},t) \zeta(\mathbf{r}',t') \rangle = -2 \kBT \omega^2 \nabla\cdot \left[M(\phi) \nabla\delta(\mathbf{r}-\mathbf{r}')\delta(t-t')\right]$ with $\omega\leq 1$ an ad hoc scaling parameter~\cite{Cook1970BrownianDecomposition,Ronsin2022PhaseFieldFilms}. The scaling parameter (artificially) dampens the magnitude of the fluctuations, allowing for larger steps in our time integrator. We have verified that this does not qualitatively change the phase-separation kinetics~\cite{Ronsin2022FormationSimulations}. Thermal fluctuations of the solution-gas interface or the (wet) film height in the lubrication equation Eq.~\eqref{eq:Lubrication} we ignore, as these are expected to be relevant only for molecularly thin films~\cite{Clarke2005TowardMixtures,Davidovitch2005SpreadingFluctuations,Grun2006Thin-FilmNoise}.

\begin{figure}[tb]
    \centering
    \includegraphics[width=\columnwidth]{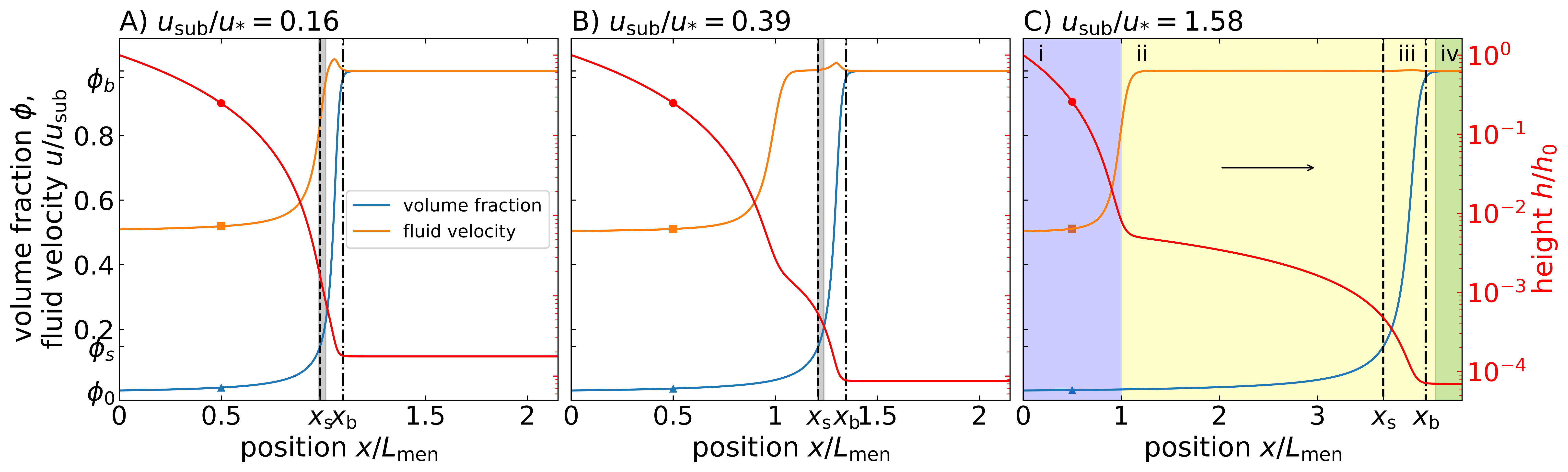}
    \caption{Calculated quasi one-dimensional calculated steady-state profiles as a function of the dimensionless position $x/L_\mathrm{men}$ for the dimensionless film height (red, circle), the volume fraction of the solute (blue, triangle) and the normalized height-averaged fluid velocity $u/u_\mathrm{sub}$ (orange, square) for a substrate velocity of A) $u_\mathrm{sub}/u_*^{} = 0.16$,  B) $u_\mathrm{sub}/u_*^{} = 0.39$ and C) $u_\mathrm{sub}/u_*^{} = 1.58$. The solution is ideal, and the other system parameters are as in Table~\ref{tab:ParameterRanges}. $L_\mathrm{men}$ is the length of the static meniscus defined in the main text. The arrow in panel C indicate the direction of the substrate motion. The vertical dashed and dashed-dotted lines denote the positions of where the solution concentration crosses the (low concentration) spinodal (at $x_\mathrm{s}$) and (high concentration) binodal (at $x_\mathrm{b}$), for the non-ideal solution conditions of Table~\ref{tab:ParameterRanges}. The gray regions in A) and B) indicate the spatial extent of the computational domain in the $x$-direction presented in the quasi two-dimensional calculations of section~\ref{sec:twodim}. We annotate four different regions of space in panel C: (i) the static meniscus in purple,  (ii) the drying front in yellow, (iii) the immiscible region between the dashed and dash-dotted line and (iv) the dry film in green.}
    \label{fig:SimulationDomain}
\end{figure}

Our model is a generalization of the models frequently used for studying the coating of binary solutions~\cite{Doumenc2010DryingPhases,LeBerre2009FromThickness,Wedershoven2018PolymerConvection}. For ideal fluids with $\chi = 0$ and (supercritical) non-ideal fluids with $\chi < 2$, our model reproduces the steady-state deposition, which we illustrate in Fig.~\ref{fig:SimulationDomain} using quasi one-dimensional calculations, where we presume that all state variables are independent of the transverse $y$ coordinate. The three cases are for different coating regimes as shown in Fig.~\ref{fig:Evap_LandauLevich}: the evaporative regime with $u_\mathrm{sub}/u_*^{}= 0.16$ in \ref{fig:SimulationDomain}A, the transition zone from the evaporative to the Landau-Levich with $u_\mathrm{sub}/u_*^{}= 0.39$ in \ref{fig:SimulationDomain}B, and the Landau-Levich regime with $u_\mathrm{sub}/u_*^{}=1.58$ in \ref{fig:SimulationDomain}C. The volume fraction is shown in blue (triangle), the fluid velocity in orange (square), and the height in red (circle) on a logarithmic scale. The background color in Fig.~\ref{fig:SimulationDomain}C we return to below. The position $x$ is scaled to the length of the static meniscus, defined as $L_\mathrm{men} = \sqrt{d h_0/2}$ with $d/2$ the curvature of the static meniscus between two parallel plates separated by a distance $d$ and $h_0$ the height of the meniscus at $x/L_\mathrm{men} = 0$ that is a somewhat arbitrarily chosen parameter, which must be the height of the film somewhere within the so-called static meniscus, see also the discussion near Fig.~\ref{fig:experimental_setup}, where the small slope condition $\nabla h \ll 1$ holds. In Fig.~\ref{fig:SimulationDomain}A the distance $d$ is the distance between the substrate and the coating device shown. Highlighted in the figure are the concentrations at the inlet $\phi_0$, the low concentration spinodal $\phi_\mathrm{s}$ at position $x_\mathrm{s}$ (vertical dashed line) and the high concentration binodal $\phi_\mathrm{b}$ at position $x_\mathrm{b}$ (vertical dash-dotted line) of the subcritical non-ideal solution discussed in the next section. The deposition of an ideal fluid that we show in Fig.~\ref{fig:SimulationDomain} serves both as reference state and as the initial condition for calculations of phase-separating fluids. Hence, it seems appropriate to discuss Fig.~\ref{fig:SimulationDomain} in detail.

We can, for our purposes, distinguish four regions in the steady-state profiles in Fig.~\ref{fig:SimulationDomain}, which we have colored in Fig.~\ref{fig:SimulationDomain}C for clarity: (i) $x/L_\mathrm{men} \leq 1$ in purple, (ii) $x/L_\mathrm{men} > 1$ up to the point where the film is dry and becomes flat in yellow, (iii) the region in between the spinodal and binodal positions of the equivalent subcritical solution, between the dashed and dash-dotted lines, that may overlap with the previous region. Finally, we have the region (iv) in Fig.~\ref{fig:SimulationDomain}C highlighted in green where the film is flat and dry. The first region (i) for $x/L_\mathrm{men} \leq 1$, indicates the width of a static meniscus, where the height profile of the solution layer (red) is dictated by the capillary pressure. The solute concentration shown in blue remains low and the fluid velocity deviates from unity due to the capillary pressure. In the second region, (ii), the film dries and the profile locally depends on the substrate velocity. The differences in the height profiles of Fig.~\ref{fig:SimulationDomain}A-C originate from the different amounts of solvent that evaporate on the time scale of withdrawal of the film. In Fig.~\ref{fig:SimulationDomain}A, solvent evaporation is relatively rapid and the solute dries at a position just beyond the static meniscus where the volume fraction rapidly increases to unity. In Fig.~\ref{fig:SimulationDomain}B, however, evaporation of the solvent is relatively slow and a wet film is deposited that subsequently dries, as is to be expected upon approaching the Landau-Levich regime. The ``shoulder'' of the height profile is indicative for the transition from the evaporative to the Landau-Levich regime, where solvent evaporation, capillary forces and viscous forces all play a r\^ole. For even higher velocities, the film dries much further removed from the meniscus as is shown in Fig.~\ref{fig:SimulationDomain}C, which is to be expected in the Landau-Levich regime. The ``bump'' in the velocity profile near the drying front, as seen in panels A and B, originates from gradients in the disjoining pressure. This maximum decreases in magnitude with increasing substrate velocity as the width of the drying front increases with increasing substrate velocity. 

The third region (iii) along the slab is that where for the subcritical solution the immiscible region is situated between the position $x_\mathrm{s}$, where it crosses the spinodal (dashed line) and $x_\mathrm{b}$, where it crosses the binodal (dashed-dotted line). See also the path the solution traverses through the phase diagram indicated in Fig.~\ref{fig:Phase_Diagram}. Only in the immiscible region can we distinguish between ideal or supercritical non-ideal mixtures and mixtures that phase separate. The reason why we focus attention in this work in particular on the unstable region near the low concentration branch of the spinodal, instead of the binodal, is that phase separation during the casting of an organic thin films typically only occurs for concentrations larger than the low concentration spinodal on account of the slowness of nucleation~\cite{Franeker2015cosolvents,Franeker2015spincoating}. The positions of both edges of the immiscible region depend on the Flory interaction parameter $\chi$, noting that domains that are advectively transported beyond the binodal position redissolve. Redissolution is, of course, a property of a binary solution and in a sense an artifact of our simplified model, whereas typically mixtures of multiple incompatible solutes in a common solvent are used that usually remain phase-separated after the solvent has evaporated and the solutes have solidified~\cite{Goffri2006MulticomponentThreshold,Peng2023ASemiconductors}. 

Finally, the fourth and final region (iv) refers to positions beyond the drying front where the film is dry and consists only of solute molecules. Here, our model reproduces the well-known relation between the dry film height and the substrate velocity shown in Fig.~\ref{fig:Evap_LandauLevich}, and discussed in the introduction. We reiterate that for low substrate velocity in the evaporative regime (nearly) all solvent evaporates in the meniscus region, which results in the dry-film height being inversely proportional to the substrate velocity~\cite{Doumenc2010DryingPhases,LeBerre2009FromThickness}. At high substrate velocity in the Landau-Levich regime, however, solvent evaporation in the meniscus is negligible on the timescale of the withdrawal of the liquid film. Hence, a wet film is in that case deposited, which subsequently dries due to solvent evaporation~\cite{Landau1988DraggingPlate,Doumenc2010DryingPhases,LeBerre2009FromThickness}. Due to a balance of the viscous and capillary forces, the dry film height scales as $h_\mathrm{dry} \sim u_\mathrm{sub}^{2/3}$~\cite{Landau1988DraggingPlate}. As discussed by Le Berre \textit{et al.}~\cite{LeBerre2009FromThickness}, the critical velocity $u_*$ that separate these two regimes is related to the material properties of the fluid and the evaporation rate as $u_* \sim F^{3/5} \sigma^{2/5}/ h_0 \eta^{2/5}$, with $F = \int_{0}^{L_\mathrm{men}} g(\phi) \dint x\propto k$ the integrated evaporation flux, and $k$ the mass transfer coefficient and $L_\mathrm{men}$ again the length of the static meniscus in the direction parallel to the substrate motion.

\section{Numerical Methods}\label{sec:numerics}
We solve our model numerically for ideal and subcritical non-ideal fluids in quasi two dimensions and quasi one dimension. The latter is computationally much less demanding, allowing us to calculate for longer times and for a broader range of our model parameters. The quasi one-dimensional calculations are crucial for us to study in much greater detail the phenomenology of meniscus-guided phase separation that we observe in the quasi two-dimensional calculations. We obtain the quasi one-dimensional model from the quasi two-dimensional description presented in the preceding section by assuming that all state variables are independent of the $y$-position perpendicular to the coating direction, and replacing all gradient operators by one-dimensional spatial derivatives $\nabla \to \partial/\partial x$. 

In both cases, we model meniscus coating by imposing appropriate inlet and outlet conditions as follows. First, at the inlet we fix the height $h(0,y) = h_0$ of the film and the volume fraction $\phi(0,y) = \phi_0$. The height $h_0$ is a somewhat arbitrary parameter that need not be equal to the distance $d$ between the coating head and the substrate, but needs to be the height of the solution somewhere within the static meniscus where the small slope condition $\nabla h \ll 1$ holds~\cite{Doumenc2013Self-patterningMeniscus}. Second, we equate the pressure $p$ to the Laplace pressure associated with a static meniscus between two plates separated by a distance $d$, representing the distance between the coating head and the substrate, for which the constant curvature reads $\partial^{2} h/\partial x^{2}|_{x=0} = 2/d$~\cite{LeBerre2009FromThickness}. Third, we are interested only in solutions that are miscible at the inlet, \textit{i.e.}, for which the initial concentration is below the saturation concentration and that phase separate downstream. Therefore, at the inlet we can disregard the square-gradient term in the exchange chemical potential density Eq.~\eqref{eq:FreeEnergy}, which then is a function of the inlet height and volume fraction only. The inlet boundary conditions read for all times $t$,
\begin{align}\label{eq:DirichletBC}
    h(0,y) = h_0, \quad &p(0,y) = -\sigma \dparth{h}{x}{2}\bigg|_{x=0} \equiv -2\sigma/d, \quad \\
    \phi(0,y) = \phi_0, \quad &\Delta \mu(0,y) = \frac{\kBT}{b^{3}} h_0 \dpart{f_\mathrm{loc}}{\phi}\bigg|_{\phi=\phi_0}. \nonumber
\end{align}
At the outlet positioned at $x = L \gg L_\mathrm{men}$, we impose Neumann boundary conditions,
\begin{align}\label{eq:NeumannBC}
    \dpart{h}{x}\bigg|_{x=L} = \dpart{p}{x}\bigg|_{x=L} = \dpart{\phi}{x}\bigg|_{x=L} = \dpart{\Delta \mu}{x}\bigg|_{x=L} = 0,
\end{align}
again for all times $t \geq 0$. For the quasi two-dimensional calculations, we use periodic boundary conditions in the $y$-direction, \textit{i.e.}, perpendicular to the coating direction, to eliminate edge effects associated with the finite width of the slot die. We refer to Table~\ref{tab:ParameterRanges} for the model parameter values used in this paper. The material and coating parameters are within the range of typical values for compounds and coating conditions of interest to us, but are otherwise somewhat arbitrary because we are interested in the phenomenology of the deposition process and not a direct comparison with a particular experimental result. Parenthetically, our focus on a mixture that phase separates under off-critical conditions could also help in understanding the \textit{phenomenology} that is observed in experiments on crystallizing solutes~\cite{Michels2021PredictiveCoating}. We point out that for our parameter choices the effective solute-solvent surface tension, which is related to the stiffness $\kappa$~\cite{Cahn1958FreeEnergy}, is always much smaller than the solution-gas surface tension. This is reasonable as the liquid-vapor surface tensions are typically much larger than liquid-liquid surface tensions.

\begin{table}[tb]
    \caption{List of the kinetic and thermodynamic parameters in our model and the ranges of parameters we investigate. The ``standard set'' is the set of parameters used in our calculations.}
    \label{tab:ParameterRanges}
    \centering
    \begin{tabular}{| c | c | c | c |}
        \hline
         Parameter & units & typical value & standard set \\
         \hline\hline
         $h_0$ &  \textmu m  & $\mathcal{O}\left(10^{2}\right)$ & 100 \\
         $d$ &  mm & $\mathcal{O}\left(0.1-1\right)$ & 1 \\
         $\phi_0$ &  - & $\lessapprox 0.01$ & 0.01 \\
         $\sqrt{\kappa}$ &  nm  & 1-100~\footnote{\cite{McCulloch2013PolymerScattering,Kuei2017ChainPolymers}} & 10 \\
         $\chi$ & - & $>2$ & 4 \\
         $D$ &  m$^2$ s$^{-1}$  & $\mathcal{O}\left(10^{-9} - 10^{-11}\right)$ & $10^{-10}$  \\
         $\sigma$ &  mN m$^{-1}$  & $\mathcal{O}(10^1)$~\footnote{\cite{2016CRCPhysics}} & $28$ \\
         $k$ &  \textmu m s$^{-1}$  & $\mathcal{O}\left(10^{-1} - 10^{1}\right)$ & $1.0$ \\
         $u_\mathrm{sub}$ &  \textmu m s$^{-1}$  & $\mathcal{O}\left(1 - 10^{3}\right)$~\footnote{\cite{2016CRCPhysics}} & 24, 60, 240 \\
         $\eta$ &  mPa s  &  $\mathcal{O}(0.5 - 10^{1})$~\footnote{\cite{Yaws2009ViscosityCompounds}} & 0.6 \\
         $b$ &  m  & $\mathcal{O}\left(10^{-9} - 10^{-10}\right)$ & $10^{-9}$ \\
         $A_\mathrm{H}$ &  J  & $\mathcal{O}\left(10^{-19}-10^{-21}\right)$~\footnote{\cite{Takagishi2019MethodTheory}} & $10^{-20}$ \\
         $\omega$ &  $-$  & 1 & $10^{-2}$ \\
         $u^{}_{*}$ &  \textmu m s$^{-1}$  & $\mathcal{O}\left(10^{1} - 10^{3}\right)$ & $150$ \\
         \hline
    \end{tabular}
\end{table}

We nondimensionalize our model parameters using the inlet height $h_0$ as the vertical scale and the extent of the static meniscus region $L_\mathrm{men} = \sqrt{d h_0/2}$ as the lateral scale. The ratio $\varepsilon = h_0/L_\mathrm{men}$ of the two length scales must be smaller than unity for the lubrication approximation to hold. We take the substrate velocity $u_\mathrm{sub}$ as the characteristic velocity scale and we introduce an advective time $t_0 = L_\mathrm{men}/u_\mathrm{sub}$. The relevant dimensionless numbers are the Capillary number $\mathrm{Ca} = \eta u_\mathrm{sub}/\sigma$, a what we call ``disjoining number'' $\mathrm{G} = A_\mathrm{H}\varepsilon/ 6 \pi h_0^2  u_\mathrm{sub} \eta$, which measures the strength of the Van der Waals forces relative to viscous forces, a Korteweg number $\mathrm{K} = \varepsilon^2 \kBT \kappa/b^{3} u_\mathrm{sub} \eta h_0$, which measures the strength of the solute-solvent surface tension and viscous forces, and is, in a sense, a solute-solvent Capillary number~\cite{Naraigh2010NonlinearFilms}. The Peclet number is defined as $\mathrm{Pe} = L_\mathrm{men} u_\mathrm{sub}/D$ and the evaporation number $\mathrm{E} = \varepsilon k/u_\mathrm{sub}$, measuring the evaporation velocity relative to the vertical velocity scale, which within the lubrication approximation reads $\varepsilon u_\mathrm{sub}$~\cite{Oron1997Long-scaleFilms}. See Table~\ref{tab:ParameterRangesDimless} for characteristic values for these dimensionless groups. We introduce $\Delta f =\kBT/b^{3}$ as the energy density scale. Using the dimensionless variables and operators $h \equiv h/h_0$, $\psi \equiv \psi/h_0$, $\Delta\mu \equiv \Delta\mu/\Delta f$, $\nabla \equiv L_\mathrm{men} \nabla$, $\mathbf{u} \equiv \mathbf{u}/u_\mathrm{sub}$, $t \equiv t/t_0$, $M \equiv M \Delta f/D$, $\kappa \equiv \kappa/L_\mathrm{men}^2$ (also known as the Cahn number) and $L \equiv L/L_\mathrm{men}$, our resulting nondimensionalized expressions now read
\begin{equation}\label{eq:Lubrication:dimensionless}
    \dpart{h}{t} + \nabla\cdot \left(h\mathbf{u}\right) = -E(\phi_\mathrm{max} - \phi),
\end{equation}
for the height of the film, with the dimensionless velocity
\begin{equation}
\mathbf{u} = (u_\mathrm{x},u_\mathrm{y})^{T} = (1,0)^{T} + \frac{h^2}{3}\left(\nabla \left[\frac{\varepsilon^{3}}{\mathrm{Ca}}\nabla^2 h + \frac{G}{h^3}\right] - K\frac{1}{h} \nabla \left(h |\nabla \phi|^2\right)\right),
\end{equation}
and the diffusion-advection equation
\begin{equation}\label{eq:genDifEq:dimensionless}
\mathrm{Pe}\left[\dpart{\psi}{t} + \nabla \cdot \left(\psi\mathbf{u} \right)\right] = \nabla \cdot h M \nabla \Delta \mu + \zeta,
\end{equation}
for the dimensionless solute height $\psi$ of the film. For the remainder of this text, all quantities are dimensionless. We solve Eqs.~\eqref{eq:Lubrication:dimensionless}~and~\eqref{eq:genDifEq:dimensionless} numerically using second order, central finite differences discretization in the positions and an adaptive semi-implicit Euler time integrator, implemented in parallel using PETSc~\cite{Abhyankar2018PETSc/TS:Library,Balay2024PETScPage,Balay2024PETSc/TAOManual}. Details can be found in appendix~\ref{app:numerics}. For reference, the quasi two-dimensional calculations presented in the next section take 18 days making use of 192 compute-cores on the Raven cluster of the Max Planck Computing and Data Facility.

\begin{table}[tb]
    \caption{List of the relevant dimensionless numbers and their typical values based on the typical parameter values listed in Table~\ref{tab:ParameterRanges}. The values for the dimensionless numbers in our calculations are based on Table~\ref{tab:ParameterRanges} evaluated for $u_\mathrm{sub}/u_{*}^{} = 1$ with $u_\mathrm{sub}$ the substrate velocity and $u_{*}^{}$ the critical velocity.}
    \label{tab:ParameterRangesDimless}
    \centering
    \begin{tabular}{ | c | c | c |}
        \hline
         Parameter & typical value & standard set \\
         \hline\hline
         Capillary Number Ca & $\mathcal{O}\left(10^{-7} - 10^{-3}\right)$ & $3.2\times 10^{-6} \times u_\mathrm{sub}/u_{*}^{}$\\
         Peclet Number  Pe & $\mathcal{O}\left(1 - 10^{5}\right)$ & $6.7 \times 10^{2} \times u_\mathrm{sub}/u_{*}^{}$ \\
         Korteweg Number  K & $\mathcal{O}\left(10^{-2} - 10^{2}\right)$ & $10^{-1}\times (u_\mathrm{sub}/u_*)^{-1}$ \\
         Disjoining Number  G & $\mathcal{O}\left(10^{-9} - 10^{-6}\right)$ & $1.3\times 10^{-7} \times \left(u_\mathrm{sub}/u_{*}^{}\right)^{-1}$\\
         Evaporation Number  E & $\mathcal{O}\left(10^{-3} - 10^{2}\right)$ & $1.5 \times 10^{-3} \times \left(u_\mathrm{sub}/u_{*}^{}\right)^{-1}$\\
         \hline
    \end{tabular}
\end{table}

Our numerical calculations are limited by (i) the small spatial resolution required to resolve the phase-boundaries between the solute-rich and the solvent-rich phase, and by (ii) the time required for the composition morphology to reach a steady state. These limitations make it unfeasible to solve our model on a uniform spatial grid where both the complete deposition process and the phase-separation processes can be studied simultaneously. Instead of using an approach that relies on an adaptive or non-uniform grid, we opt to tackle this issue by invoking an approach comprising three steps, which we detail below. For this, we make use of our observation that the ideal and subcritical non-ideal mixtures appear to be indistinguishable, except in the narrow regions shown in Fig.~\ref{fig:SimulationDomain} where the latter type of mixture phase-separates. Therefore, we can limit our numerical study of subcritical mixtures to this narrow region only. Moreover, the meniscus-guided deposition of an ideal or supercritical fluid turns out to be, for all intents and purposes, quasi one-dimensional. Hence, we opt to use a combination of quasi one-dimensional and quasi two-dimensional calculations to efficiently model meniscus-guided phase-separation, since the quasi one-dimensional calculations are numerically much cheaper, thus allowing for larger and longer calculations. For the same reason, we will also first use quasi two-dimensional calculations of subcritical mixtures to study the interplay of phase-separation and coating qualitatively, and subsequently use the quasi one-dimensional using a larger range in the model parameters to study this in additional detail. 

\begin{figure}[tb]
    \centering
    \includegraphics[width=0.3\columnwidth]{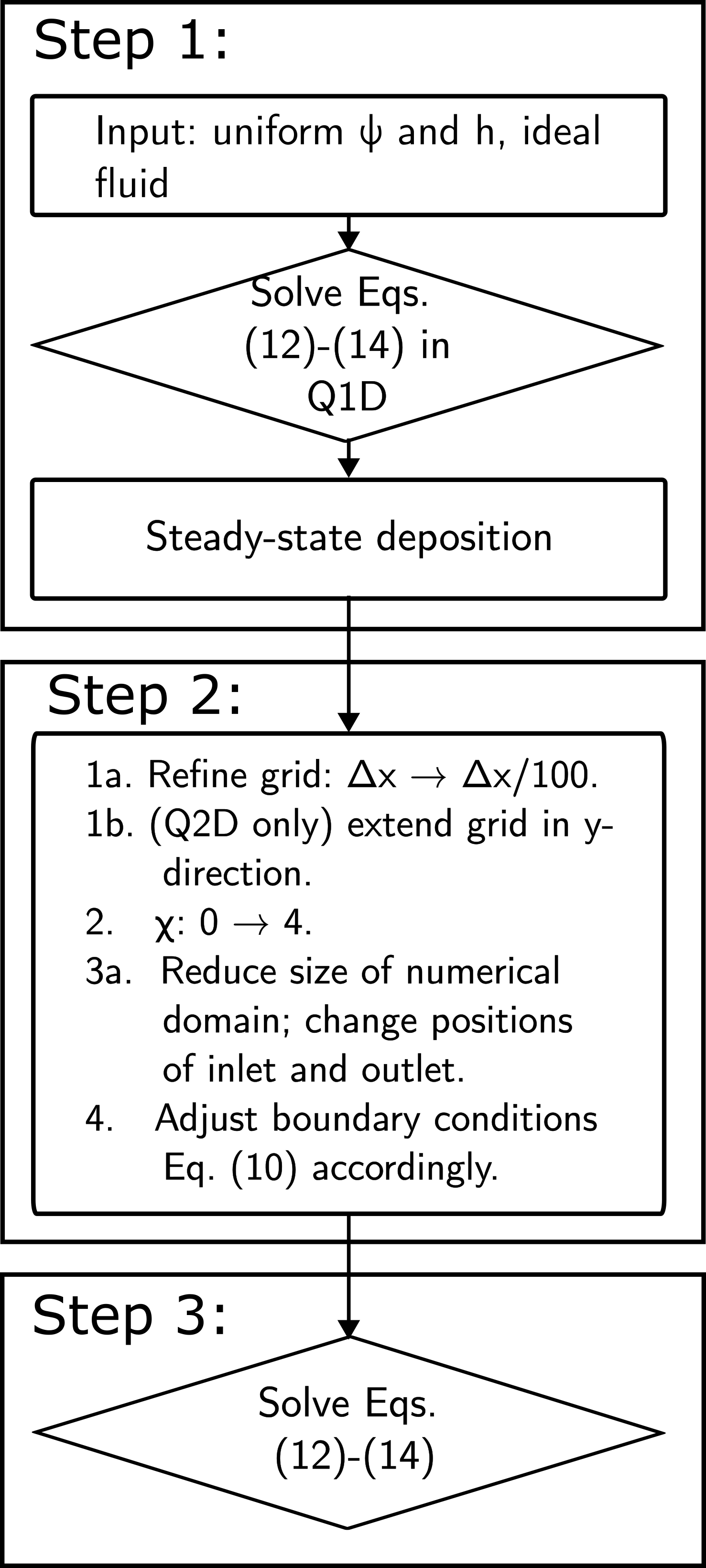}
    \caption{Flow diagram of our three step numerical approach. In step 1, we solve Eqs.~\eqref{eq:Lubrication:dimensionless}-\eqref{eq:genDifEq:dimensionless} for an ideal fluid in quasi one dimension (Q1D) until the we achieve steady state conditions. In step 2, we modify the calculation settings using the steady-state results to serve as input for our calculations on subcritical fluids. In step 3, we solve Eqs.~\eqref{eq:Lubrication:dimensionless}-\eqref{eq:genDifEq:dimensionless} in the Q1D and Q2D case for a subcritical fluid using the processed steady-state profiles from step 2. We refer to the main text for more details.}
    \label{fig:Simulation_flowchart}
\end{figure}

Before showing our results, we first explain our multi-step calculation setup, which we also show schematically in Fig~\ref{fig:Simulation_flowchart}. In the first step in our calculations, we solve the quasi one-dimensional equivalents of Eqs.~\eqref{eq:Lubrication:dimensionless} and \eqref{eq:genDifEq:dimensionless} for an ideal fluid with $\chi = 0$ on a relatively coarse grid, and fix the position of the outlet at $L= 4.5$. This is sufficiently far removed from the static meniscus to be able to model both the evaporative regime and the transition to the Landau-Levich regime, which are the regimes of interest as they provide the conditions for the desired dynamic control over the morphology in the composition~\cite{Yildiz2022OptimizedCoating}. Eqs.~\eqref{eq:Lubrication:dimensionless} and \eqref{eq:genDifEq:dimensionless} are integrated in time, starting from a uniform height and volume fraction field until we obtain steady-state conditions, which we identify with the condition that the 2-norm of the state vector $\{\psi,\Delta\mu,h,p\}$ between two consecutive time steps is smaller than $10^{-10}$. These steady-state conditions are illustrated in Fig.~\ref{fig:SimulationDomain} and discussed in Section~\ref{sec:theory} for three different coating velocities in the three different coating regimes. The steady-state configurations of an ideal fluid serve as the initial conditions for our calculations on phase-separating fluids. 

In the second step in our approach, we introduce the subcritical fluid in by instantaneously replacing at $t = 0$ the value of the Flory interaction parameter $\chi$ from zero to four, while concomitantly imposing a stiffness $\kappa$ larger than zero, thereby replacing the ideal fluid. Simultaneously, we adjust the setup of the numerical calculations to be able to resolve phase boundaries and to limit the study to the narrow region of interest where the solution can phase separate. This requires a number of steps as detailed in Fig~\ref{fig:Simulation_flowchart}. First, we refine the spatial grid and for the quasi two-dimensional calculations extend the height and volume fraction fields in the $y$-direction along 500 grid elements. We also change the positions of the ``inlet'' and outlet boundaries of the domain used in the numerical calculations. For quasi two-dimensional calculations we cannot solve for the entire immiscible region because of the very high computational cost. Instead, we solve the equations only in the gray regions indicated in Fig.~\ref{fig:SimulationDomain}A-B, where the inlet boundary is positioned 500 grid elements upstream from the position of the spinodal and the outlet 7000 grid elements downstream from it. For the quasi one-dimensional calculations, our numerical calculation domains encompass the entire immiscible region with the inlet and outlet positioned 5000 grid elements preceding and beyond the spinodal and binodal positions, respectively. Increasing the number of grid elements before the position of the spinodal does not affect our results. The values for the new inlet boundary conditions in Eq.~\eqref{eq:DirichletBC} we obtain from the calculations on the ideal fluid. Finally, in the third and final step in our approach we solve Eqs.~\eqref{eq:Lubrication:dimensionless} and \eqref{eq:genDifEq:dimensionless} for the subcritical solution. For additional details we refer to appendix~\ref{app:numerics}. 

Our approach, in which we replace an initially ideal solution by a subcritical solution effectively introduces two types of quench in our calculations: an instantaneous quench at $t=0$ for all fluid elements already in the calculation domain, as well as a finite-rate quench due to solvent evaporation for fluid elements that enter our numerical calculation domain at the inlet after $t = 0$. This also means that we expect two types of response, associated with the two different quenches. Associated with the instantaneous quench is an initial ``induction'' period before steady-state coating is reached, where the morphological evolution is dictated by this quench and not by the coating process. After the induction period, the initial conditions no longer influence phase separation, which is now only coupled to the hydrodynamic transport processes originating from solvent evaporation and coating. The situation is then that of the continuous coating of a substrate with a solution that by then has become immiscible beyond some distance from the coating device due to evaporation. 

In the next section, we study this using quasi two-dimensional calculations, focusing on the region in the film in the vicinity to the position where phase separation commences (the shaded region in Fig.~\ref{fig:SimulationDomain}A and \ref{fig:SimulationDomain}B). In the subsequent section, we study the deposition patterns we observe in the quasi two-dimensional calculations in greater detail by making use of the quasi one-dimensional calculations. We return to the initial induction period at the end of this paper.

\section{Quasi two-dimensional model}\label{sec:twodim}
Let us first focus on our quasi two-dimensional model calculation to study phase separation of a binary fluid that is deposited on a substrate via meniscus-guided deposition before delving more deeply into the phenomenology of the patterning using the computationally less demanding quasi one-dimensional calculations. Here, we limit our investigation to two substrate velocities, one in the evaporative regime ($u_\mathrm{sub}/u_* = 0.16$) and one in the transition zone between the evaporative and the Landau-Levich regime ($u_\mathrm{sub}/u_* = 0.39$), shown in Figs.~\ref{fig:SimulationDomain}A and \ref{fig:SimulationDomain}B. Movies showing the continuous deposition of the phase-separating fluids for both substrate velocities are available in the supplemental material S1 and S2. We do not present calculations deeper in the Landau-Levich regime, because these are numerically extremely costly due to the much larger spatial extent of the drying fronts. Furthermore, since the emergent phase-separation structure only very weakly affects the height and velocity fields, that is, by less than $0.1$ \% in all cases, we do not discuss these in 2D and refer to the 1D analogues presented in Fig.~\ref{fig:SimulationDomain}. 

\begin{figure}[tb!]
    \centering
    \includegraphics[width=\columnwidth]{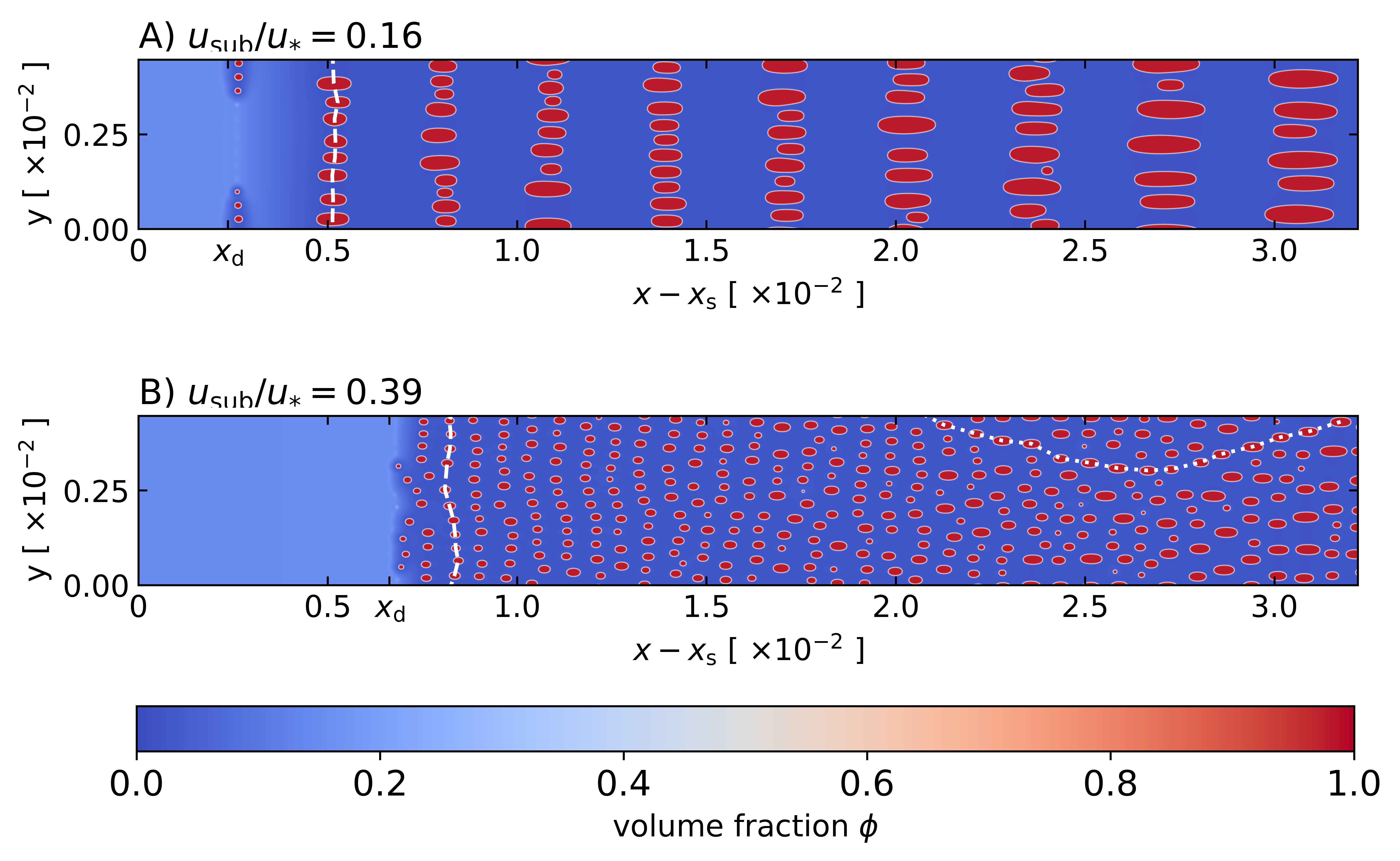}
    \caption{The volume fraction of solute as a function of the (dimensionless) position, indicated by the color coding defined at the bottom of the figure. The position is relative to that where the solute concentration has reached the low-concentration spinodal,  $x_\mathrm{s}$; $x_\mathrm{d}$ is the approximate position where the solute-rich domains form. The model parameters are given by the standard set in Table~\ref{tab:ParameterRanges}. In panel A) we show the volume fraction for coating in the evaporative regime for the substrate velocity $u_\mathrm{sub}/u_* = 0.16$ at dimensionless time $t = 0.06$, and in panel B) for coating in the transition from the evaporative to the Landau-Levich regime for $u_\mathrm{sub}/u_* = 0.39$ at dimensionless time $t = 0.08$. The color coding is indicated at the bottom of the figure. The white vertical dashed lines indicate the center-of-mass positions in an array of droplets. The white dotted line indicates one of the horizontal strings of seemingly correlated droplets. The relevant dimensionless parameters are (A) $K = 6\cdot 10^{-1}$, Ca = $5\cdot 10^{-7}$, Pe = $53$, G = $1.6 \cdot 10^{-6}$, and (B) $K = 2.6\cdot 10^{-1}$, Ca = $1\cdot 10^{-6}$, Pe = $134$, G =$6.6\cdot 10^{-7}$. The substrate movement is from left to right. The calculation grid shown measures 3600 by 500 grid elements. See also Movies S1 and S2.}
    \label{fig:2DFinalTime}
\end{figure}

In Fig.~\ref{fig:2DFinalTime} we show snapshots of our numerical calculations for the volume fraction for values of the model parameters of our standard set quoted in Table~\ref{tab:ParameterRanges}, for the two substrate velocities under investigation, $u_\mathrm{sub}/u_* = 0.16$ (Fig.~\ref{fig:2DFinalTime}A) and $u_\mathrm{sub}/u_* = 0.39$ (Fig.~\ref{fig:2DFinalTime}B). The substrate moves from left to right. We reiterate that we solve our model equations only in the regions shaded in gray in Fig.~\ref{fig:SimulationDomain}A and \ref{fig:SimulationDomain}B. The snapshots are limited to 3600 grid elements in the $x$-direction for clarity, which is about half of the 7500 grid elements in the calculations. To distinguish between the two cases shown in Fig.~\ref{fig:2DFinalTime}, we refer to the top conditions (with $u_\mathrm{sub}/u_* = 0.16$) as slow coating and the bottom conditions (with $u_\mathrm{sub}/u_* = 0.39$) as fast coating. For both the slow- and fast-coating cases, phase separation commences at some distance downstream from the position of the low volume-fraction spinodal, $x = x_\mathrm{s}$, and forms an array of droplet-like, solute-rich domains in a solvent-rich continuous phase. In case of slow coating these domains form (nearly) straight lines perpendicular to the coating direction, whereas for fast coating, this ordering is lost. The droplet-like structure is characteristic for off-critical spinodal decomposition, as is to be expected from the gradual quench in the solute concentration due to solvent evaporation~\cite{Rogers1989NumericalSeparation}.

For the low coating velocity, represented by Fig.~\ref{fig:2DFinalTime}A, the arrays of solute-rich domains are deposited periodically, and the position of the centers of mass of each domain within an array lie on an (almost) straight line perpendicular to the coating direction, indicated with the vertical white dashed line in Fig.~\ref{fig:2DFinalTime}A. The average number of domains that form per unit of dimensionless length in each line is between $22 \cdot 10^{2}$ and $25 \cdot 10^{2}$, which we find by counting for 17 arrays the number of domains that form divided by the size of our calculation grid in the $y$-direction. After the initial deposition of the droplets, the domains grow anisometrically due to the combined effects of (i) extensional flow caused by the curvature of solution-gas interface and (ii)  Ostwald ripening. In our case, Ostwald ripening is anisotropic due to the broken translational and orientational symmetry and is therefore distinct from the typical situation of Ostwald ripening in a spatially homogeneous material~\cite{Lifshitz1961TheSolutions,Wagner1961TheorieOstwaldReifung}.

For the case of fast coating, shown in Fig.~\ref{fig:2DFinalTime}B, demixing commences at a position further removed from the spinodal, which we explain below. The deposition pattern still appears to be periodic in the $x$-direction, but the droplets are now being deposited in ``wavy'' rather than straight lines as seen for slow coating (Fig.~\ref{fig:2DFinalTime}A). The number of domains per unit of length in each curved line is slightly larger than during slow coating: about $25-29 \cdot 10^{2}$, which we determine based on the deposition of 50 arrays of droplets. This is to be expected as the contour length of the (imaginary) curve that connects the centers of mass of the deposited droplets within a wavy line, illustrated in Fig.~\ref{fig:2DFinalTime}B with the vertical dashed white line, is larger than for the straight arrays of Fig.~\ref{fig:2DFinalTime}A. Furthermore, for fast coating the increased disorder in the deposition morphology suggests that the effect of thermal fluctuations is larger than in the slow coating case. This seemingly counter-intuitive observation stems from the difference in the mechanism that determines the morphological evolution for slow versus fast coating that we discuss in the next section. Both the wavelength of the (quasi) periodic deposition and the size of the solute-rich domains are significantly smaller compared to the slow coating case, but only in the $x$-direction parallel to the direction of coating. The periodicity in the $x$-direction itself becomes less prominent for later times (further downstream from the inlet) due to ripening and coarsening: with increasing distance from the inlet or, equivalently, age of the domains, the structural morphology becomes increasingly more isotropic and the initial periodic, straight structure becomes increasingly less discernible. 

Some subtle patterns may, however, remain, as we observe in Fig.~\ref{fig:2DFinalTime}B that the centers-of-masses of the larger domains appear to form strings of droplets along the $x$-direction, weakly aligned with the fluid flow. We indicate a single string of droplets with the white dotted line in Fig.~\ref{fig:2DFinalTime}B. This is reminiscent of the string-like structures that form when spherical colloidal particles are subjected to shear flow~\cite{Cheng2012AssemblyFlow,SantosdeOliveira2012TheFluids}. Since the flow field (not shown) is not strongly affected by phase separation itself, presumably because in our model the viscosity is independent of the composition and solute-solvent interfacial tension is weak in comparison with the solution-gas surface tension, we conclude that the origin of these string-like structures is not caused by the distortion of flow field. As it turns out, the largest domains in subsequent lines tend to form at a $y$ position only slightly deviating from the previous line. These larger domains tend to resist Ostwald ripening, which is a diffusive coarsening mechanism that promotes the growth of large domains at the expense of small ones. This can in fact be observed from the movie S2 provided in the supplemental material. Apparently, coarsening destroys the periodic patterning perpendicular to the coating direction, but can expose subtle patterns parallel to  the coating direction. Since the coarsening kinetics are outside of the scope of this paper, we do not study this in more detail. Instead, we focus next on the initial stages of phase separation. 

The initial stages of phase separation are dominated by the amplification of the most unstable (spinodal) density mode until phase separation occurs. 
With this amplification we can associate the spinodal lag time $\tau_\mathrm{L}$, which quantifies the delay in time between crossing the spinodal and the unstable spinodal density modes to sufficiently amplify, and a emergent wave number that sets the relevant phase-separation length scale just after demixing. Both these quantities depend non-trivially on the quench conditions, which include the rate of evaporation~\cite{Binder1983CollectiveMixtures,Huston1966SpinodalCooling,Schaefer2016StructuringEvaporation}. If the differences in the deposition structure and the characteristic length of solute-rich domains shown in Fig.~\ref{fig:2DFinalTime} originate from the coupling of the hydrodynamic transport processes to the initial stages of phase separation, this coupling must result in a change in the spinodal lag time $\tau_\mathrm{L}$ and the emerging wave number $q_*$. For meniscus-guided deposition, the lag time can be translated into a what we call a demixing position $x_\mathrm{d}$ where the solute-rich domains form, which depends on the substrate velocity as Fig.~\ref{fig:2DFinalTime} demonstrates. 

Because the fluid velocity between the positions $x_\mathrm{s}$ and $x_\mathrm{d}$ deviates only by about $4$ \% for slow coating and $10^{-3}$ \% for fast coating, we deduce that the position $x_\mathrm{d} \approx u_x \tau_\mathrm{L}$, with $u_x$ the $x$-component of the velocity field evaluated at the position $x_\mathrm{s}$ where the volume fraction crosses the spinodal. This shows that the initial phase separation is not significantly affected by the meniscus-guided deposition. Indeed, the \textit{local} fluid velocities at the position of the spinodal, which need not be equal to the substrate velocity as shown in Fig.~\ref{fig:SimulationDomain}, are for slow and fast coating $0.11$ and $0.39$, respectively, in agreement with the ratio of the demixing positions $x_\mathrm{d}$ of approximately three. This suggests that the lag time $\tau_\mathrm{L}$ is indeed independent of the coating velocity and essentially set by the process of spinodal decomposition. 

We have actually verified this by analyzing a linearized version of Eq.~\eqref{eq:genDifEq}, assuming that the spatial variations in the fluid velocity between the spinodal and the lag positions can be neglected. From this analysis, detailed in Appendix.~\ref{app:earlytime}, we conclude that the initial amplification is identical to that in a stationary and flat thin film albeit driven by solvent evaporation, and that the lag time therefore depends only on the kinetic and thermodynamic properties of the fluid mixture and on the solvent evaporation rate. In other words, the coupling of the deposition and demixing processes in flow is weak because of the homogeneous viscosity (one of the assumptions in our model).

Basing ourselves on the same analysis, we also conclude that the emergent wave number $q_*$, a measure for the characteristic feature size at the position $x_\mathrm{d}$, is independent of the coating velocity. Indeed,
\begin{equation}\label{eq:critical_wavenumber}
    q_* \approx |\partial^3 f_\mathrm{loc}/\partial \phi^3\alpha|^{1/2} \tau_\mathrm{L}^{1/2},
\end{equation}
where $\alpha = \mathrm{E} \phi(x_\mathrm{s})/h(x_\mathrm{s}) (\phi_\mathrm{max} - \phi(x_\mathrm{s}))$ is a measure for the evaporation rate, with $\mathrm{E}$ the evaporation number and $\phi_\mathrm{max}$ a cut-off in our evaporation model, introduced in Section~\ref{sec:theory}~\cite{Schaefer2015StructuringEvaporation}. Both the evaporation rate and the third derivative of the local free energy are evaluated at the position of the spinodal and therefore constant, referring to Appendix~\ref{app:earlytime} and also to Schaefer \textit{et al.}~\cite{Schaefer2015StructuringEvaporation}. Note that the predictions of Eq.~\eqref{eq:critical_wavenumber} deviate from that of the classical case of spinodal decomposition after an instantaneous quench, because we presume in our analysis a non-zero evaporation rate and that we cross the low concentration branch of the spinodal at $t = 0$. Hence, Eq.~\eqref{eq:critical_wavenumber} only holds for non-vanishing evaporation rates. From Eq.~\eqref{eq:critical_wavenumber} for the emergent wave number, we must conclude that the hydrodynamic transport processes for meniscus coating and in particular the substrate velocity do not affect the initial stages of the phase separation kinetics. Hence, the differences in the emergent length scale and structure that we find in Fig.~\ref{fig:2DFinalTime}, in particular the increase in the domain size and the distance between subsequently deposited arrays of domains with decreasing velocity, must originate some form of interplay between the hydrodynamic transport and the (diffusive) growth of the domains after demixing. 

In order to investigate this in greater detail, in the next section we introduce a quasi one-dimensional representation of our model that is computationally much less demanding and also easier to analyze. This enables us to study a larger range of the model parameters, as well as larger numerical calculation domains along the coating direction. As we show below, the wavelength of the periodic deposition and the initial domain size are intrinsically connected, and both can be understood from a competition in the mass transport processes in the depletion zones around solute-rich domains.

\section{Periodic deposition: quasi one-dimensional model}\label{sec:onedim}
The quasi two-dimensional calculations discussed in the preceding section show that the coating velocity has a considerable influence on the characteristic compositional length scales and morphology. In this section, we first focus on the periodic deposition at low substrate velocities in order to study the correlation between the phase separation length scale and coating velocity. For computational convenience we make use of the quasi one-dimensional versions of Eqs.~\eqref{eq:Lubrication:dimensionless}~and~\eqref{eq:genDifEq:dimensionless}, again assuming that the state variables only depend on the $x$ position and time $t$, and replace the gradient operators in Eqs.~\eqref{eq:Lubrication:dimensionless}~and~\eqref{eq:genDifEq:dimensionless} by one-dimensional spatial derivatives $\nabla \to \partial/\partial x$. This we justify by the observation that the results for the height-averaged, quasi two-dimensional results shown in Fig.~\ref{fig:2DFinalTime} show a (nearly) one-dimensional deposition structure as the domains are deposited on (nearly) straight lines that run perpendicular to the coating direction. The subsequent late-stage aging and coarsening of the domain structure, which appear to destroy the periodic pattern in Fig.~\ref{fig:2DFinalTime}B, is obviously not captured within a quasi one-dimensional model~\cite{Lifshitz1961TheSolutions,Wagner1961TheorieOstwaldReifung}. The numerical domains for the quasi one-dimensional calculations span the full immiscible region in Fig.~\ref{fig:SimulationDomain}A-C, as discussed in Section~\ref{sec:numerics}.

\begin{figure}[tb]
    \centering
    \includegraphics[width=0.95\textwidth]{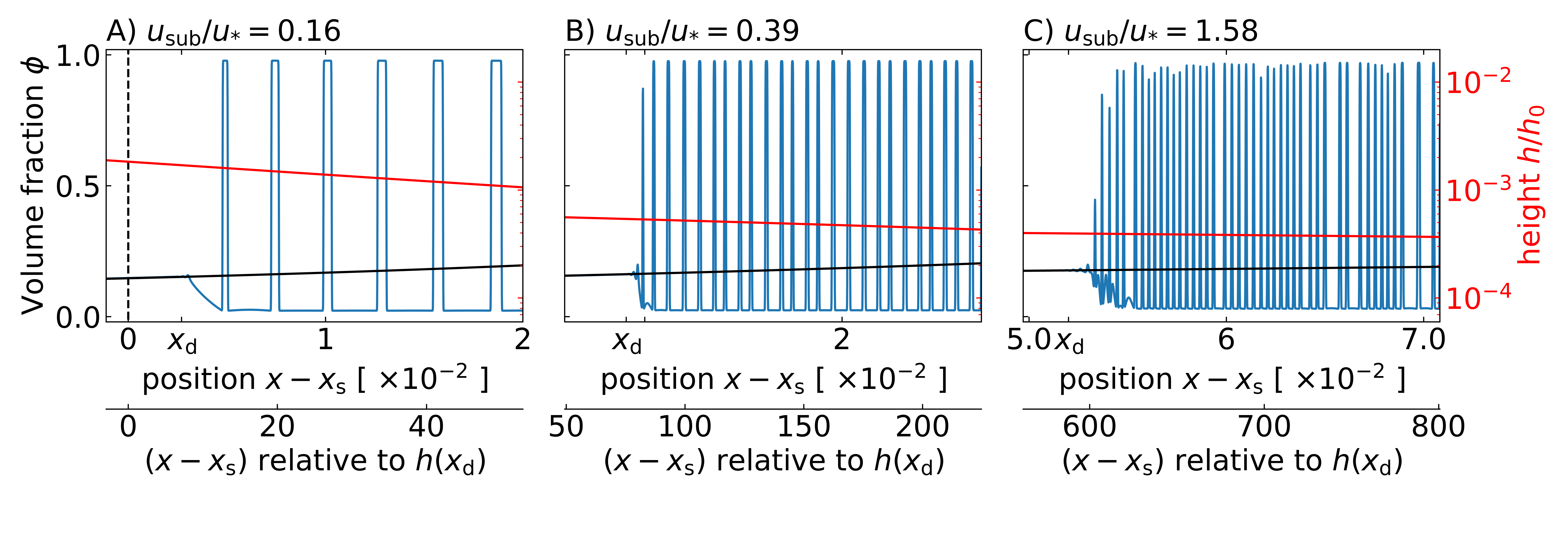}
    \caption{The local volume fraction $\phi$ after demixing of a subcritical solution in blue, the volume fraction of an ideal solution in black and the height of the solution-gas interface in red as a function of the position, measured starting from the position $x_\mathrm{s}$ where the lower spinodal is crossed. Lower x-axis: the position divided by the corresponding height $h$ of the film at the position where the solution phase separates $x_\mathrm{d}$. Coating conditions are those of our standard set, detailed in Table~\ref{tab:ParameterRanges}. A): Substrate velocity equal to $u_\mathrm{sub}/u_*^{} = 0.16$ at time $t = 4$; start-up effects in our calculations disappear after $t = 0.1$. B): Substrate velocity equal to $u_\mathrm{sub}/u_*^{}=0.39$ at time $t = 3.9$; start-up effects disappear after $t = 0.1$, C): Substrate velocity equal to $u_\mathrm{sub}/u_*^{}=1.58$ at time $t = 1.5$; start-up effects disappear after $t = 0.3$. The relevant dimensionless parameters are (A) $K = 6\cdot 10^{-1}$, Ca = $5\cdot 10^{-7}$, Pe = $53$, G = $1.6 \cdot 10^{-6}$, (B) $K = 2.6\cdot 10^{-1}$, Ca = $1\cdot 10^{-6}$, Pe = $134$, G =$6.6\cdot 10^{-7}$, (C) $K = 6.3\cdot 10^{-2}$, Ca = $5\cdot 10^{-6}$, Pe = $530$, G =$1.6\cdot 10^{-7}$.}
    \label{fig:1DSnapshots}
\end{figure}

The crossover from periodic to random deposition of solute-rich domains survives in the (quasi) one-dimensional model, as Fig.~\ref{fig:1DSnapshots} shows. Indicated in Fig.~\ref{fig:1DSnapshots} are snapshots of the film height (red, top), the solute volume fraction for the subcritical binary solution (blue, sawtooth-shaped curve) with the same parameter values as those used for Fig.~\ref{fig:2DFinalTime} and the volume fraction for the steady-state coating of an ideal solution (black, bottom) as a function of the (dimensionless) position. The latter quantity acts as a measure for the mean solute concentration at a given position. The coating velocities $u_\mathrm{sub}/u_* = $ 0.16 and $u_\mathrm{sub}/u_* = $ 0.39 for \ref{fig:1DSnapshots}A and \ref{fig:1DSnapshots}B are identical to the coating velocities in the quasi two-dimensional calculations shown in Fig.~\ref{fig:2DFinalTime}A-B. The coating velocity in \ref{fig:1DSnapshots}C is $u_\mathrm{sub}/u_* = 1.58$, solidly within the Landau-Levich regime. Again, $x_\mathrm{s}$ is the position where the volume fraction crosses the volume fraction of the spinodal and $x_\mathrm{d}$ is the estimated position where the solution demixes. The secondary horizontal axis (below the primary one) re-expresses the position $x - x_\mathrm{s}$ scaled to the height of the film at $x_\mathrm{d}$. From this, we deduce that in all three cases the lateral size of the domains just after their inception is slightly larger than the height of the film. This is true for all our other calculations also. Note, however, that we consider a rather deep quench with Flory interaction parameter $\chi = 4$, whereas experimentally relevant quenches are much more shallow for which the expected feature size is much larger. Hence, our vertically averaged model should yield a reasonable description of experimental reality.

In Fig.~\ref{fig:1DSnapshots}A, we notice the periodic deposition of relatively large domains that grow in size with position or, equivalently, age. This growth originates from the redistribution of solute and solvent material due to solvent evaporation. Solvent evaporation causes the volume fractions in the solute-rich and solute-poor phases to reach values beyond those of the binodals, which results in a diffusive redistribution of the molecules. At the same time, non-uniform evaporation decreases the height of the film but does so more rapidly in the solvent-rich than in the solute-rich phase. The capillary pressure induced by the curved gas-liquid interface counteracts this, resulting in a hydrodynamic redistribution of the solute domains over a larger area. The periodic deposition turns out to be almost deterministic and the deposition wavelength that measures the center-of-mass distance between consecutive domains is nearly invariant of position. This agrees with our quasi two-dimensional calculations at the same coating velocity, as is shown in the Fig.~\ref{fig:2DFinalTime}A.

For coating in the transition region between the evaporative and the Landau-Levich regime, shown in Fig.~\ref{fig:1DSnapshots}B, we obtain a similar trend as in our quasi two-dimensional calculations of Fig.~\ref{fig:2DFinalTime}B. The domains are smaller than for the slower coating velocity of Fig.~\ref{fig:1DSnapshots}A and the distance between consecutive domains fluctuates weakly, again in agreement with our finding that the effect of thermal fluctuations becomes more pronounced at higher coating velocities. The domains do not visually increase in size with position over the range shown. The reason is that because of the higher substrate velocity the age of the domains increases much less across the same spatial extent. Incidentally, this is also evident by the small slope of the height of the film (red, top) when compared to the Fig.~\ref{fig:1DSnapshots}A. 

Finally, in the Landau-Levich regime shown in Fig.~\ref{fig:1DSnapshots}C for $u_\mathrm{sub}/u_*^{}= 1.58$, we find a slightly different behavior with multiple domains forming in rapid succession between $x_\mathrm{d}$ and $x-x_\mathrm{s} \approx 5.5 \times 10^{-2}$, before absorbing all solute molecules from the neighboring solute-poor phase. Because these domains compete from the same solute reservoir, this may result in insufficient material for the volume fraction in all domains to approach the equilibrium (binodal) value. The unstable domains eventually disappear, resulting in the distance between subsequent domains to double, arguably introducing a secondary depositions length scale.

\begin{figure}[bt]
    \centering
    \includegraphics[width=0.75\columnwidth]{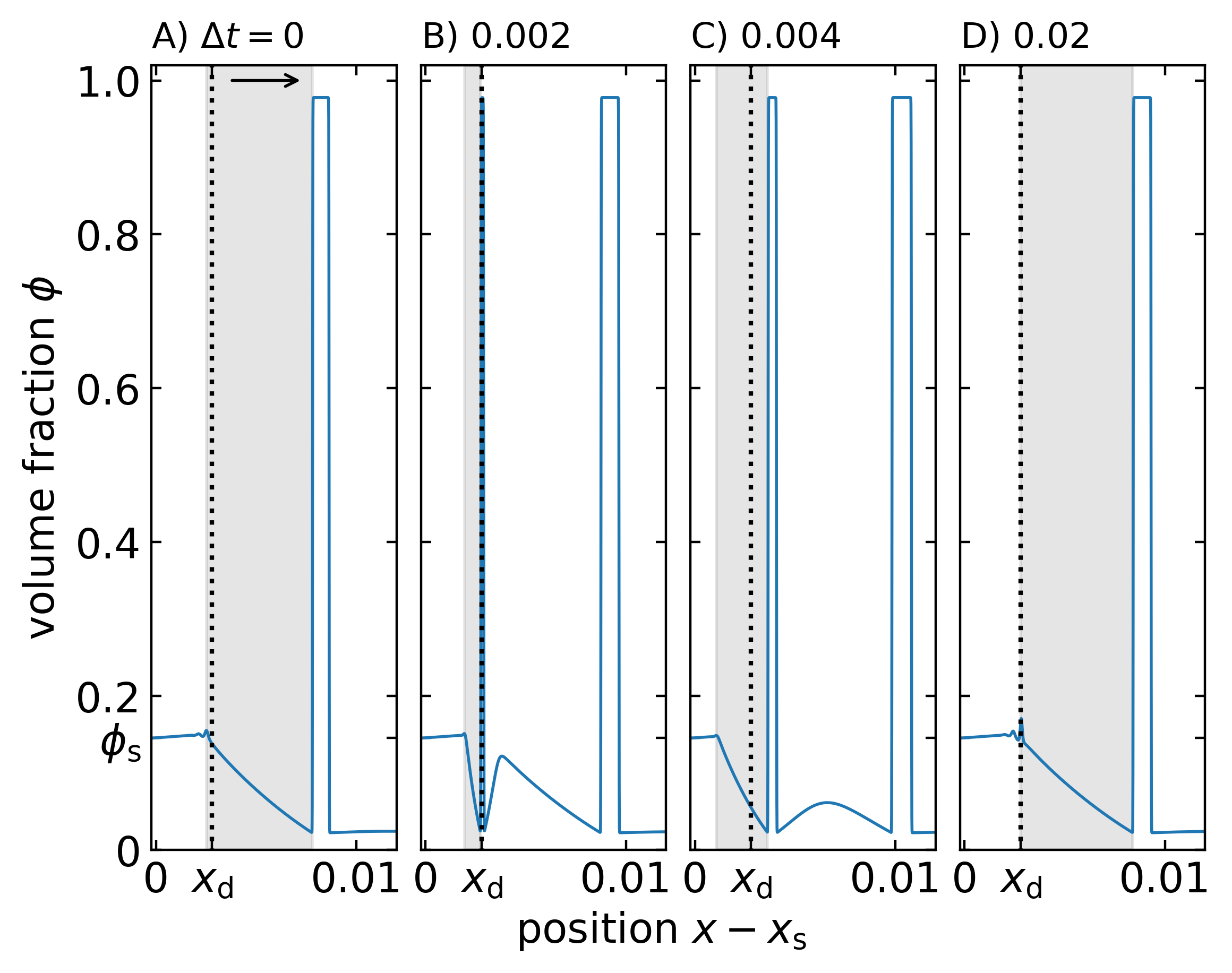}
    \caption{The volume fraction $\phi$ in blue as a function of the dimensionless position relative to the spinodal position $x - x_\mathrm{s}$ at different moments in time for the substrate velocity $u_\mathrm{sub}/u_*^{} = 0.11$ and the other kinetic and thermodynamic parameters identical to our standard set in Table~\ref{tab:ParameterRanges}. A) The time $\Delta t$ relative to snapshot A, which itself is in our calculation at the dimensionless time $t = 9.34$, with the start-up effects disappearing at $t = 0.15$. In panel B) snapshot at time $\Delta t = 2\times 10^{-3}$, C) $\Delta t = 4\times 10^{-3}$ and D) $\Delta t = 2\times 10^{-2}$. Vertical dotted line is the estimated demixing position $x_\mathrm{d}$. The arrow in panel A indicates the direction of motion. The regions shaded in gray are a guide for the eye to indicate the extent of the depletion zone.}
    \label{fig:PeriodicDeposition}
\end{figure}

Before delving into how the emergent length scale depends on the coating velocity, we first discuss in detail the physical origin for the periodic deposition of the solute-rich domains for slow coating shown in Fig.~\ref{fig:1DSnapshots}A. The periodic deposition, as it turns out, plays a crucial role in determining the emergent compositional length scale. To understand this, we focus on the solute-rich domain facing the meniscus, as shown in Fig.~\ref{fig:PeriodicDeposition} for a substrate velocity of $u_\mathrm{sub}/u_*^{} = 0.11$. This is somewhat lower than the substrate velocity $u_\mathrm{sub}/u_*^{} = 0.16$ used in Figs.~\ref{fig:2DFinalTime}A and ~\ref{fig:1DSnapshots}A, because the effect we focus on magnifies with decreasing velocity. Again, the domains form at a position $x_\mathrm{d}$, on account of the delay between crossing the spinodal position and the actual moment in time that the mixture phase separates. This position is indicated with the vertical dotted line. The regions shaded in gray highlight the size of the solute depletion zone upstream from the solute-rich domain closest to the position of the spinodal. 

At some time $\Delta t=0$ shown in Fig.~\ref{fig:PeriodicDeposition}A, the edge of the depletion zone coincides with the demixing position. The oscillation in the volume fraction near $x = x_\mathrm{d}$ stems from the spinodal instability that grows until the fluid (locally) phase separates. Indeed, at or slightly after this moment in time, a new domain forms as Fig.~\ref{fig:PeriodicDeposition}B shows. This new domain is accompanied by new depletion zones and inhibits the growth of the depletion zone of the previously deposited domain. In the downstream direction (left to right), the depletion zone merges with the (upstream) depletion zone of the nearest solute-rich domain. In the upstream direction, the depletion zone can grow freely and may extend to positions closer to the spinodal than the demixing position $x_\mathrm{d}$ as is illustrated in Fig.~\ref{fig:PeriodicDeposition}C. This temporarily inhibits the formation of a new domain and results in an increased growth time for the solute-rich domain closest to the demixing position, and consequently, in a larger depletion zone, a larger solute-rich domain and an increased deposition wavelength than the length scale of the spinodal instability. Finally, due to the advective motion of the fluid (from left to right), the edge of the depletion zone returns to the demixing position, shown in Fig.~\ref{fig:PeriodicDeposition}D. Here, we observe the same spinodal instability as in Fig.~\ref{fig:PeriodicDeposition}A that precedes in the formation of a new domain albeit that the detailed structure is obviously different due to the presence of noise. This process repeats periodically resulting for slow substrate velocities in the periodic deposition of domains (much) larger than the length scale of the spinodal instability. 

Based on all of this, we conclude that the maximum size of the depletion zone upstream of the solute-rich domain closest to the position of the spinodal $x_\mathrm{s}$ sets the compositional length scale under slow-coating conditions. This is a depletion length scale that we attribute to a competition between two types of mass transport. First, new material is continuously deposited on the substrate from the meniscus, supplying new solute material to the film. Second, solute molecules diffuse towards the solute-rich domains, resulting in depletion zones around all domains.  Implicit, but imperative to this competition, is the directionality of the coating, as the periodic deposition of solute-rich domains relies on two counteracting transport processes. Based on a simple Peclet number argument, relying on the fluid velocity at the demixing position $u_x(x_\mathrm{d})$ parallel to the direction of coating and the diffusion coefficient $D$ as characteristic measures for the two counteracting transport processes, we conclude that the depletion length scale must be proportional to $D/u_x(x_\mathrm{d})$. This depletion length scale is the size of the depletion zone shown in Fig~\ref{fig:PeriodicDeposition}D. The reason why we must use the local fluid velocity at the demixing position $u_x(x_\mathrm{d})$ instead of the substrate velocity, is that the former describes the \textit{local} hydrodynamic transport processes relevant for demixing that may deviate in magnitude from the substrate velocity as we illustrated in Fig.~\ref{fig:SimulationDomain}.

Obviously, in our depletion argument we implicitly assume that the depletion length is much larger than the initial size of the domains, while the latter is in fact proportional to the spinodal length $L_\mathrm{spin} = 2 \pi /q_*$, with $q_*$ the critical wave number Eq.~\eqref{eq:critical_wavenumber} as argued in section~\ref{sec:twodim}. Hence, the depletion argument can only be valid if the spinodal length is much smaller than the depletion length. If, on the other hand, the depletion length is smaller than the spinodal length, which is the case for the results shown in Fig.~\ref{fig:1DSnapshots}C at higher substrate velocities, the latter sets the emergent compositional length scale. An additional naive argument why this is to be expected, is that at sufficiently high substrate velocity, deep in the Landau-Levich regime, the deposited film is flat and stationary with respect to the substrate. It subsequently dries over a distance proportional to the substrate velocity. Hence, for high coating velocities, phase separation commences far removed from the meniscus under conditions similar to that in a stationary flat thin film~\cite{Schaefer2016StructuringEvaporation,Negi2018SimulatingInvestigation}. 

In the light of this, it turns out to be instructive to introduce the spinodal Peclet number $\Pe_\mathrm{spin} = u_x(x_\mathrm{d}) L_\mathrm{spin}/D$ as the characteristic measure for the mechanism that sets the emergent morphology. For $\Pe_\mathrm{spin} \ll 1$ the depletion length is much larger than the spinodal length, and we therefore expect that the emergent length $\langle L \rangle$ scales as 
\begin{equation}\label{eq:periodic_deposition:slow_velocity_scale}
\langle L \rangle \propto \Pe_\mathrm{spin}^{-1}\sim u_x(x_\mathrm{d})^{-1}.
\end{equation}
For $\Pe_\mathrm{spin} \gg 1$ spinodal decomposition sets the emergent morphology that in that case is independent of the coating velocity and $\langle L \rangle \propto L_\mathrm{spin}$. This actually also explains why the compositional morphology shown in Figs.~\ref{fig:2DFinalTime} and \ref{fig:1DSnapshots} appears to be affected more by thermal fluctuations during fast coating than during slow coating. In the former case, the morphological evolution is dictated by the initial stages of spinodal decomposition which is known to depend sensitively on the thermal noise, whereas in the latter case, depletion is what sets the morphological evolution, which is a process that is nearly independent of thermal noise ~\cite{Cook1970BrownianDecomposition,Schaefer2015StructuringEvaporation}.

Now that we understand the scaling of the emergent compositional length in the limits of very fast and very slow coating, we next study in detail how it depends on intermediate values of the coating velocity using numerical calculations. In order to quantify the phase separation length scales in our quasi one-dimensional calculations, we introduce a single time-averaged mean length that we calculate from the dynamic structure factor as
\begin{equation}\label{eq:defmeanL}
    \langle L \rangle = \frac{2 \pi}{\langle q \rangle} = 2 \pi\left(\frac{1}{\Delta t}\int_{t'}^{t' + \Delta t} \dint t \left\{\frac{\int S(q,t) q \dint q}{\int S(q,t) \dint q}\right\}\right)^{-1}.
\end{equation}
Here, the dynamic structure factor $S(q,t) = \langle|\widehat{\delta \phi}(q,t)|\rangle^2$ is the Fourier transform of the position-dependent volume fractions and the angular brackets denote the ensemble average, where $\widehat{\delta \phi}(q,t) = \widehat{\phi}(q,t) - \widehat{\phi}(q,0)$. The positions of the lower spinodal $x_\mathrm{s}$ and the upper binodal $x_\mathrm{b}$, illustrated in Fig.~\ref{fig:SimulationDomain}, are the integration boundaries for the Fourier transform. The integration boundaries for the integrals over the wave number in Eq.~\eqref{eq:defmeanL} are $2 \pi/(x_\mathrm{b} - x_\mathrm{s})$ and $2 \pi /\Delta x$, with $\Delta x$ the grid spacing. We average over a time window $\Delta t$, which is at least the time required for the domains that form at the demixing position $x_\mathrm{d}$ at $t = t'$ to be redissolved at $t = t' + \Delta t$. The offset time $t'$ is the moment in time where the start-up features no longer affect our results, and where steady state is reached. Both these quantities depend on the coating velocity because the distance between the positions of the spinodal and the binodal increases with increasing coating velocity. Although multiple length scales associated with the domain size or the deposition wavelength can be identified in Figs.~\ref{fig:2DFinalTime}~and~\ref{fig:1DSnapshots}, the deposition wavelength is the dominant one as it remains approximately constant and is therefore associated with Eq.~\eqref{eq:defmeanL}.

\begin{figure}[bt]
    \centering
    \includegraphics[width=\textwidth]{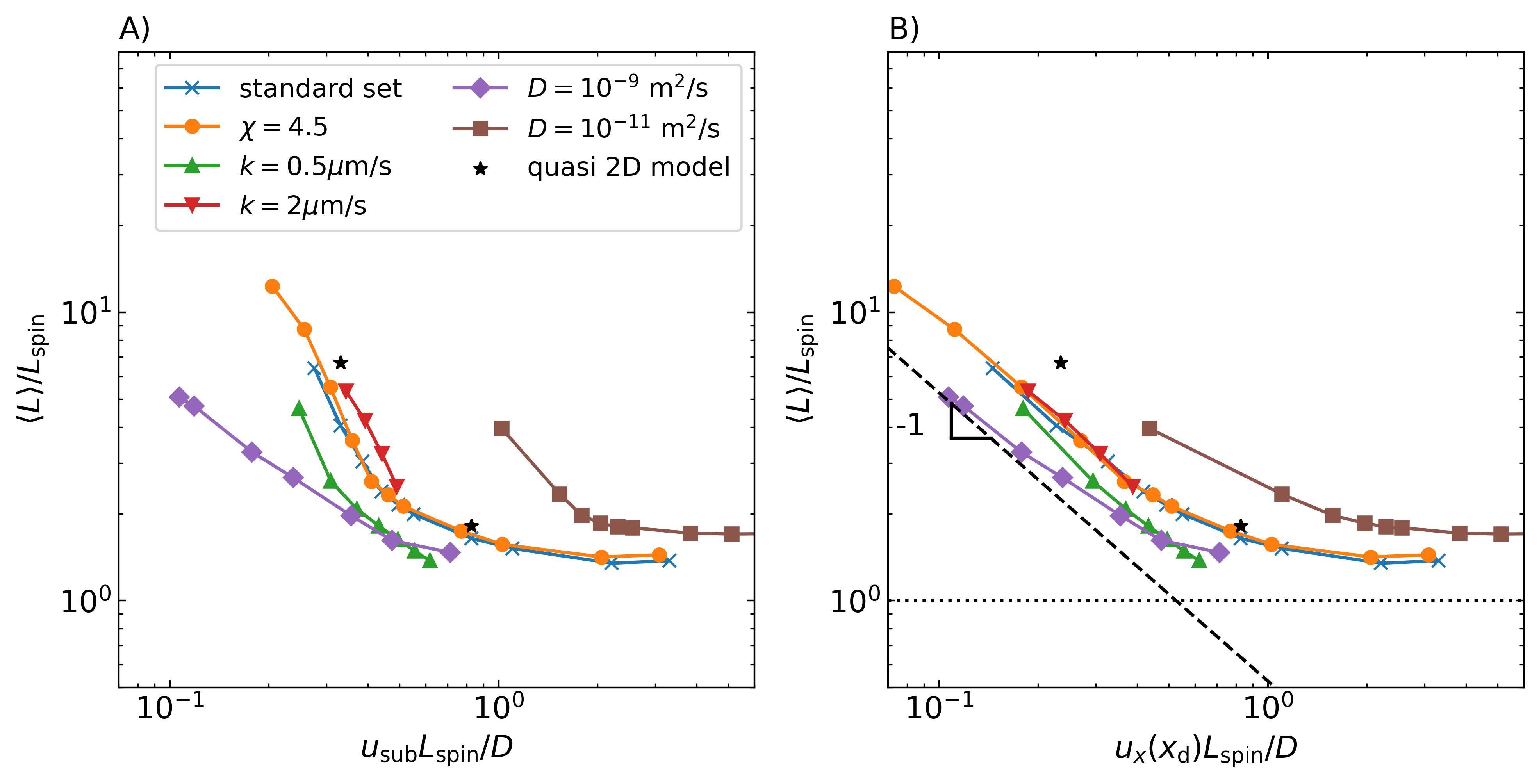}
    \caption{A) The mean length $\langle L \rangle$ scaled to the spinodal length scale $L_\mathrm{spin}$ as a function of the substrate velocity divided by the diffusivity $D$. The data for calculations with the model parameter values given by our standard set of Table~\ref{tab:ParameterRanges} are shown in blue (crosses). All other colors are for calculations with parameters of our standard set where we change a single parameter. In orange (disks) the Flory interaction parameter is $\chi = 4.5$, in green (triangle-up) the evaporative mass transfer coefficient $k = 0.5$ \textmu m/s, in red (triangle-down) $k = 2$ \textmu m/s, in purple (diamonds) the tracer diffusivity $D = 10^{-9}$ m$^{2}$/s and in brown (squares) $D = 10^{-11}$ m${^2}$/s. For the standard set, these parameter values are set at $k = 1$ \textmu m/s, $D = 10^{-10}$ m${^2}$/s and $\chi = 4.0$. The black stars are  estimates for the deposition wavelength of the quasi two-dimensional calculations shown in Fig.~\ref{fig:2DFinalTime}. B) Same data as in panel A, now as a function of the spinodal Peclet number, see main text. The dashed lines are a guide for the eye indicating the expected scaling $\langle L \rangle \propto 1/u_x(x_\mathrm{d})$. See also Eq.~\eqref{eq:periodic_deposition:slow_velocity_scale} and the main text.}
    \label{fig:MeanLengthScaled}
\end{figure}

In Fig.~\ref{fig:MeanLengthScaled} we show for a variety of thermodynamic and kinetic parameters the normalized mean length $\langle L \rangle/L_\mathrm{spin}$ as a function of two characteristic velocities, non-dimensionalized by multiplication by the spinodal length scale $L_\mathrm{spin}$ and division by the diffusion constant $D$. In Fig.~\ref{fig:MeanLengthScaled}A we use the substrate velocity $u_\mathrm{sub}$, as it is an actual control parameter in an experimental setting. In Fig.~\ref{fig:MeanLengthScaled}B we instead take the fluid velocity at the demixing position $u_x(x_\mathrm{d})$, and we may identify the $x$-axis as the spinodal Peclet number defined above. The spinodal length scale we estimate as $L_\mathrm{spin} \equiv 2 \pi/q_*$, as discussed in the previous section; see Eq.~\eqref{eq:critical_wavenumber}. The black stars are the estimated deposition wavelengths for the quasi two-dimensional model shown in Fig.~\ref{fig:2DFinalTime} that we extract by measuring (in real space) the distances between the consecutive array of droplets in the downstream direction. For the sets of parameters studied in Fig.~\ref{fig:MeanLengthScaled}, we change in each case a single parameter in our standard set in Table~\ref{tab:ParameterRanges}: the quench depth $\chi: 4 \to 4.5$ in Fig.~\ref{fig:MeanLengthScaled} shown in orange, the evaporation rate via the mass transfer coefficient $k =1 \textmu m/s$ to $k = 2$ \textmu m/s in red and $k = 0.5$ \textmu m/s in green and the tracer diffusivity from $D = 10^{-10}$ m$^{-2}$/s to $D = 10^{-9}$ m$^{-2}$/s in purple and $D = 10^{-11}$ m$^{-2}$/s in brown.

As we can conclude from Fig.~\ref{fig:MeanLengthScaled}A, varying the parameters of our model does not result in qualitative changes, only in quantitative changes. For instance, varying the diffusivity $D$ leads to shifts of the spinodal and depletion lengths, as well as the position where the solution phase separates. The reason is that the spinodal lag time, and therefore the demixing position, depends on the diffusivity. The local fluid velocity at this different demixing position may also differ, and this leads to different predominant length scales, at least when plotted as as a function of the substrate velocity. Similarly, changing the solute-solvent interaction strength $\chi$ changes the spinodal length and the position where the solution demixes, because $\chi$ controls the quench depth and therefore the concentration at the spinodal, which also affects the position of the spinodal $x_\mathrm{s}$. Finally, the evaporative mass transfer coefficient $k$ not only affects the phase separation length scale, as can be deduced from Eq.~\eqref{eq:critical_wavenumber}, but also influences the deposition process, which shifts the critical velocity separating the evaporative and the Landau-Levich regimes. Irrespective of the specific values of the thermodynamic and kinetic parameters, we find a transition from a (near) constant deposition length to a rapidly increasing deposition length with decreasing substrate velocity. These turn out to be the asymptotic regimes for the mean length in the slow and fast coating limit, respectively.

What is clear from Fig.~\ref{fig:MeanLengthScaled}A is that $u_\mathrm{sub}$ cannot be the appropriate velocity scale. So, in Fig.~\ref{fig:MeanLengthScaled}B we plot the data shown in Fig.~\ref{fig:MeanLengthScaled}A as a function of the spinodal Peclet number, based on the velocity $u_x(x_\mathrm{d})$. The data more or less collapses onto the same curve. This suggests that the spinodal Peclet number is indeed the appropriate parameter that controls the morphological transition, and that different values of the model parameters can be understood as changing the depletion or the spinodal length, or both. Basing ourselves on these results, we may conclude that kinetic control over the for the application relevant length scale is only possible if the substrate moves sufficiently slowly, on the order of the growth rate of solute-rich domains. Precisely how slowly the substrate should move depends on the diffusivity of the solute and the spinodal length that is controlled by the evaporation rate, the diffusivity of the solute (and solvent) and arguably also on the gradient stiffness $\kappa$, although we have not studied this in Fig.~\ref{fig:MeanLengthScaled}. In practice, this quantity depends on the interaction strength between the solute and solvent $\chi$ and the molecular weight of the solute ~\cite{deGennes1980DynamicsBlends,Debye1959AngularMixtures}. Interestingly, the morphological transition occurs for substrate velocities below that of the transition from the evaporative to the Landau-Levich regime for the parameter space that we  analyzed, and our explanation for the transition also does not rely on the presence of this transition. Hence, we find it not to be coupled to either the evaporative or Landau-Levich regimes, and we therefore conclude that kinetic control over the emergent phase-separation length scale is possible irrespective of the coating regime.

\section{Induction period}\label{sec:initialtransient}
The dynamic steady-state deposition of a subcritical binary solution that we studied in Sections~\ref{sec:twodim} and \ref{sec:onedim} is in practice always preceded by a transient, non-steady regime. During this period of time, which we refer to as the induction period, start-up effects are important and the deposited structure need not be identical to that during the steady-state conditions. In general, understanding the processes taking place during the start-up period is important from an application point of view, as these determine the time interval prior to the actual steady state production stage. This interval is ideally kept as short as possible to minimize material waste. 

In our numerical calculations we expect at least two different types of start-up effect due to our multi-step numerical approach detailed in Section~\ref{sec:numerics}. The first we associate with the initial step in our calculations, where we numerically solve for the deposition of an ideal fluid starting from homogeneous initial conditions implying a flat deposition layer with a height equal to $h_0$ until we reach steady-state conditions. Since the initial homogeneous state is used only out of numerical convenience, we deem this start-up effect physically irrelevant. The second type of start-up effect we associate with the subsequent replacement of this ideal fluid by one that is subcritical, which introduces a relaxation from a well-mixed state toward the (steady-state deposition of a) demixed solution. As also discussed in Section~\ref{sec:numerics}, this second type of start-up effect can be understood in terms of the response of the mixture to an instantaneous quench from an ideal or supercritical state to a subcritical state. In this section, we study this second start-up effect, because it may give rise to phase separation behavior that is different from that during the steady-state deposition regime. In the following, we first study the phenomenology of phase-separation during the induction period by making use of quasi two-dimensional calculations. Next, we discuss the induction time and how it depends on the coating conditions. Finally, we use quasi one-dimensional calculations to show that the time evolution of the compositional morphology during the induction period is, in a sense, universal.

\begin{figure}[tb]
    \centering
    \includegraphics[width=\textwidth]{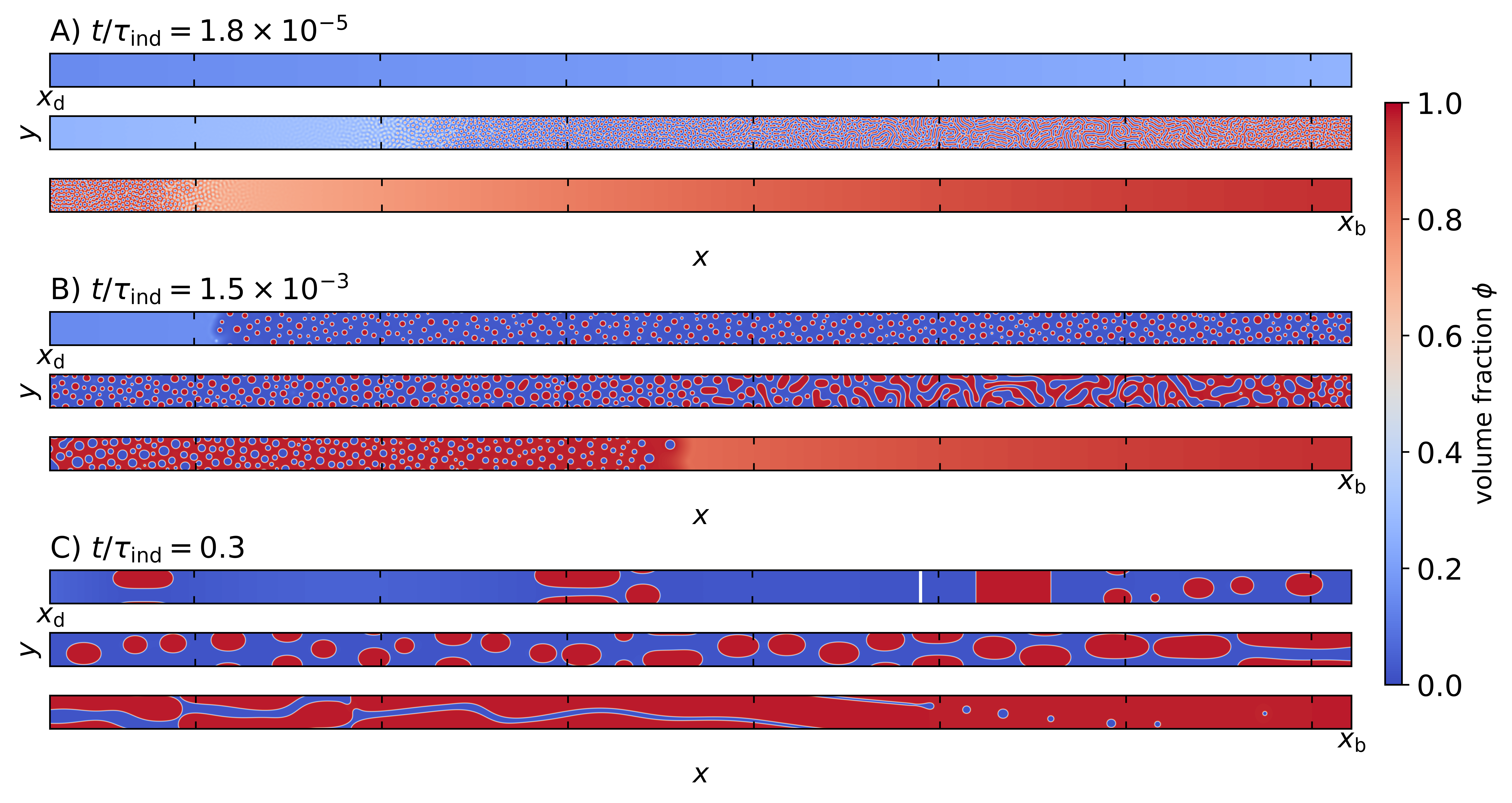}
    \caption{Snapshots of the response to the initial instantaneous quench on a narrow numerical domain (12500 by 100 grid points) at three different times. The snapshots are subdivided in the downstream direction into 3 sub-snapshots each spanning $3.5 \times 10^{-2}$ in dimensionless length, shown vertically with increasing position from top to bottom. After a short delay time during which the solution remains mixed, (A) phase separation commences in a small region in the immiscible region, which then (B) quickly spreads throughout the immiscible region. (C) Due to the motion of the substrate (left-to-right) the domains are advected towards the outlet, and new domains form near the spinodal demixing position. Indicated times $t/\tau_\mathrm{ind}$ are normalized to the time $\tau_\mathrm{ind}$ required for a domain that forms at the demixing position $x_\mathrm{d}$, located at the left boundary of the top panels, to redissolve at the redissolution position $x_\mathrm{b}$, located at the right edge of the bottom panels. The white line in (C) represents the distance that a fluid parcel that was initially at the demixing position $x_\mathrm{d}$ has traversed. Model parameters are $\Ca = 2.1 \times 10^{-7}$, $\Pe = 4.5\times 10^{2}$, $\mathrm{K} = 3\times 10^{-1}$, $\mathrm{G} = 2.0 \times 10^{-6}$ and $\mathrm{E} = 2 \times 10^{-2}$. The images shown are part of a movie presented in the supplemental material, S3.}
    \label{fig:2D_early-to-late}
\end{figure}

In Fig.~\ref{fig:2D_early-to-late} we show three snapshots at different moments in time during the induction period for the quasi two-dimensional case on a numerical domain of 12500 by 100 grid elements, and we include the movie S3, on which this figure is based, in the supplementary material. The colors encode the solute fraction, interpolating between red (high value) to blue (low value). The height field is not shown, because it remains (nearly) unaffected by the demixing of the solution. The model parameters are given by the Capillary number $\Ca = 2.1 \times 10^{-7}$, the Peclet number $\Pe = 4.5\times 10^{2}$, the Korteweg number $\mathrm{K} = 3\times 10^{-1}$, the disjoining number $\mathrm{G} = 2.0 \times 10^{-6}$ and the evaporation number $\mathrm{E} = 2 \times 10^{-2}$. These values deviate from those of standard set in order be able to model the complete immiscible region, that is, the region spanned by the low concentration spinodal and the high concentration binodal, at a reasonable computational cost. Again, we need not consider the regions of space where the solution is metastable on account of the sluggishness of nucleation. The left boundary of the top slice is the demixing position $x_\mathrm{d}$ where the domains form during the steady-state meniscus-guided deposition discussed in the previous sections. The position of the high concentration binodal is located at the right boundary of the bottom panels. The time is scaled to the induction time $\tau_\mathrm{ind}$ that we define below. The snapshots are divided along the $x$-direction into 3 slices each spanning a dimensionless length of $3.5 \times 10^{-2}$ and stacked vertically for clarity. 

After a short delay in time after the instantaneous quench at $t = 0$, the solution starts to phase separate as shown in Fig.~\ref{fig:2D_early-to-late}A. Phase separation turns out to initiate at the position where the solute volume fraction equals approximately one-half, $\phi \approx 1/2$, and subsequently rapidly spreads throughout the region of space where the volume fraction is between that of the spinodals as shown in Fig.~\ref{fig:2D_early-to-late}B. When the phase-separation fronts approach the positions of either the low or high concentration spinodal, phase separation slows down and eventually stops before actually reaching them. This is the expected behavior, based on our finding that these early stages of demixing occur on time scales much shorter than those associated with hydrodynamic transport and solvent evaporation. Due to this separation of time scales, the (initial) phase separation kinetics are decoupled from the deposition process. Moreover, at $t = 0$ the volume fraction field varies over length scales much larger than the expected spinodal feature size. Hence, we can treat each fluid element of the solution as locally homogeneous and phase separation is then dictated by the local conditions only, in this case the \textit{local} volume fraction at the moment of the instantaneous quench at $t=0$~\cite{Binder1983CollectiveMixtures,Cahn1958FreeEnergy,Cahn1959FreeFluid}. Indeed, we find that phase separation initiates at the position where the spinodal lag time, so the delay in time required for the unstable density modes to amplify, is the shortest. The theoretically predicted spinodal lag time for a homogeneous solution after an instantaneous quench and in absence of evaporation reads $4 \kappa \Pe  \frac{\partial^2f_\mathrm{loc}}{\partial \phi^2}$, with $\kappa$ the dimensionless interfacial stiffness and $f_\mathrm{loc}$ the dimensionless local free energy density given in Eq.~\eqref{eq:locFreeEnergyDenst}~\cite{Binder1983CollectiveMixtures}. This spinodal lag time is minimal for $\phi = 1/2$~\cite{Binder1983CollectiveMixtures} and diverges upon approaching both the low- and high-concentration spinodals. 

On top of the different moments in time that the fluid phase separates locally, the emergent morphology also changes in the downstream direction (left to right) from a dispersion of solute-rich droplets in a solvent-rich continuous phase via a bicontinuous structure into an inverted arrangement comprising a dispersion of solvent-rich droplets in a solute-rich matrix. Moreover, we conclude from Fig.~\ref{fig:2D_early-to-late}A and \ref{fig:2D_early-to-late}B that the typical feature size increases with age of the domains and with the degree of off-criticality, \textit{i.e.}, for volume fractions increasingly smaller or larger than one-half. The former shows that phase separation and ripening occur simultaneously during these early stages albeit that these processes happen at different positions. The latter is in agreement with our assessment that the local conditions set the phase-separation kinetics: we find that the emergent wave number $q_*$, which characterizes the typical domain size just after demixing, attains its maximum value at $\phi = 1/2$ and decreases to zero at the volume fraction of the spinodals~\cite{Binder1983CollectiveMixtures}. In other words, spinodal demixing starts at the position where $\phi = 1/2$ and from there spreads throughout the region spanned by the spinodals. The typical feature size of the phase-separated domains is smallest at $\phi = 1/2$ and increases with the degree of off-criticality. Demixing does not appear to occur for positions where the solution is metastable of where spinodal time scale is large, \textit{i.e.}, on the order of the hydrodynamic or evaporative time scale, as in that case the evaporation-assisted phase separation mechanism discussed in the previous sections becomes the dominant demixing mechanism. All this is in agreement with what we observe in Figs.~\ref{fig:2D_early-to-late}A and \ref{fig:2D_early-to-late}B and movie S3 in the supplemental material.

The above discussion is only accurate for the early stages of demixing, when the separation of demixing and evaporation and hydrodynamic time scales holds. Since that the spinodal lag time diverges for positions close to the spinodals, the demixing process eventually slows down and then occurs on time scales longer than those for hydrodynamic transport and solvent evaporation. Hence, we find a crossover to a second demixing regime wherein hydrodynamics and evaporation then start to influence the demixing and coarsening dynamics. Furthermore, the advective motion (from left-to-right) results in the supply of new material into the film and the redissolution of the phase-separated mixture near the position of the high-concentration binodal as shown in Fig.~\ref{fig:2D_early-to-late}C (right edge of bottom slice). The new material that is supplied to the film also becomes unstable and phase separates but does so via the evaporation-assisted phase-separation mechanism discussed in the preceding sections. These domains form at the (steady-state) demixing position, positioned at the left edge of the top sub-snapshots. Hence, we expect that for later times in the induction period the phase-separated morphology is determined by a combination of meniscus-guided phase separation and phase separation in response to an instantaneous quench. These two mechanism act in two distinct regions: one where the domains form as a response to the instantaneous quench at $t = 0$ (downstream of the white line shown in Fig.~\ref{fig:2D_early-to-late}C) and one where the domains form via the meniscus-guided phase separation (between the steady-state demixing position and the white line shown in Fig.~\ref{fig:2D_early-to-late}C). The (white) line that separates these two regions moves to the outlet at a rate equal to the local fluid velocity; the fluid elements at this white line were still at the (steady state) demixing positions in Fig.~\ref{fig:2D_early-to-late}A and \ref{fig:2D_early-to-late}B. The elongated droplet structures downstream of the white line originates from a combination of coarsening, solvent evaporation and extensional flow due to the curvature of the solution-gas interface. On account of the interplay of all these effects, the morphological evolution is highly complicated and a plethora of phenomena can be found. For example, we may observe what appears to be a Rayleigh-Plateau instability in the bottom two slices in Fig.~\ref{fig:2D_early-to-late}C, where droplets are pinched off from a fluid filament or ``jet'' of the solvent-rich phase; see also the movie S3 in the supplemental material~\cite{deGennes2004CapillarityPhenomena}. Evidently, the late-time morphological evolution is far from trivial. We return to this at the end of this section.

\begin{figure}[tb]
    \centering
    \includegraphics[width=0.5\textwidth]{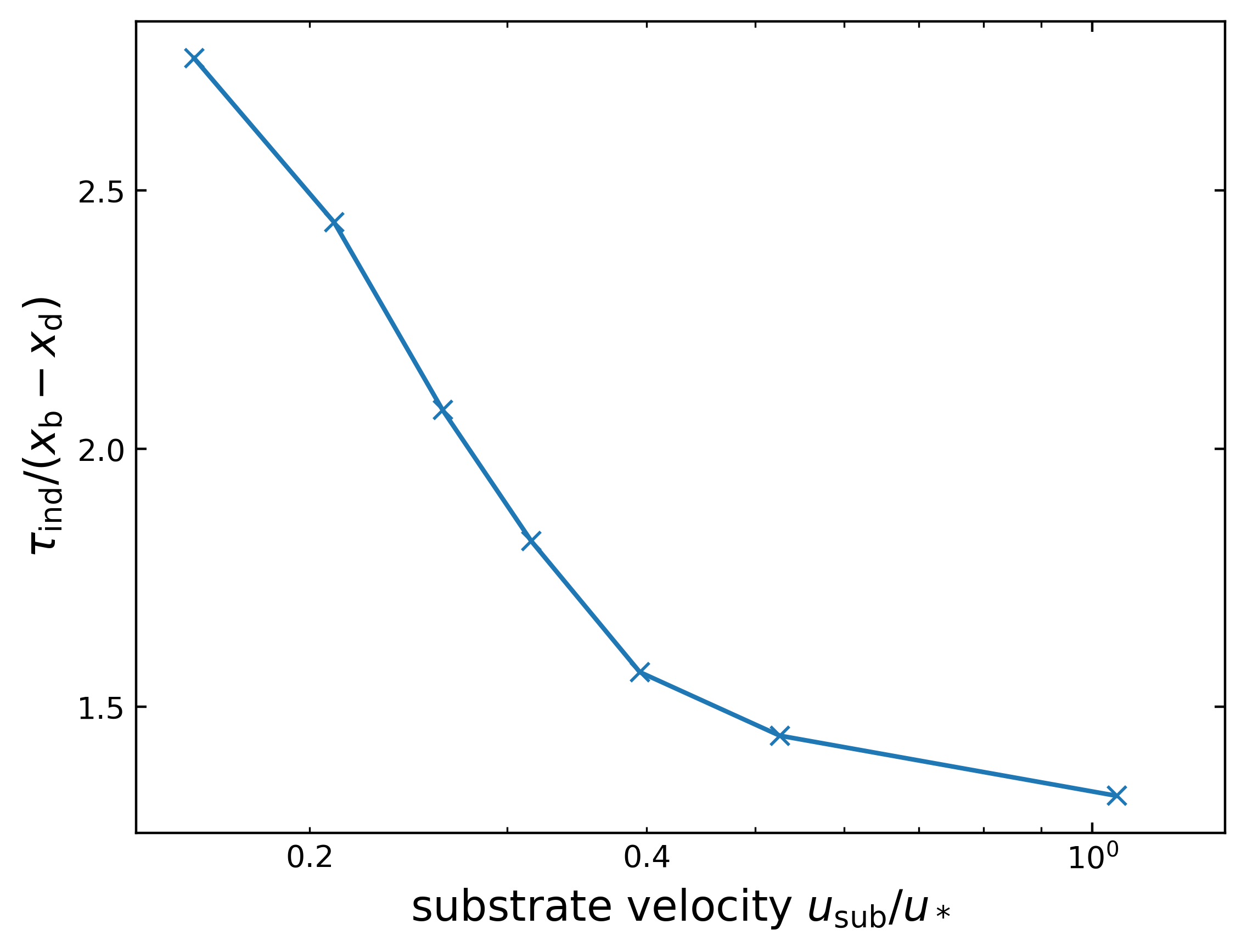}
    \caption{The (dimensionless) induction time scaled to the (dimensionless) distance between positions of the high-concentration binodal $x_\mathrm{b}$ and the position where the solution demixes $x_\mathrm{d}$, as function of the scaled substrate velocity $u_\mathrm{sub}/u_*$. Here, $u_*$ is the critical velocity that separates the evaporative and Landau-Levich regimes. Conditions of the calculations are those of the standard set in Table~\ref{tab:ParameterRanges}.}
    \label{fig:inductiontime}
\end{figure}

Now that we qualitatively understand the morphological evolution during the induction period, we next return to the duration of this period. Taking our observations in Fig.~\ref{fig:2D_early-to-late} as inspiration, we argue that start-up effects dissipate when the white line crosses the position of the high concentration branch of the binodal, for then all domains that form as a response to the instantaneous quench have redissolved. Again, redissolution is an artifact of our simple two-component mixture. In an experimental context, the position of redissolution may arguably be replaced by that of solidification. Hence, we put forward that the induction period must be proportional to the time required for a fluid parcel that at $ t = 0$ finds itself at the demixing position to be advectively transported to the position where the domains redissolve $x_\mathrm{b}$. In other words, $\tau_\mathrm{ind} \approx \int_{x_\mathrm{d}}^{x_\mathrm{b}} u_x^{-1}(x) \dint x \propto x_\mathrm{b} - x_\mathrm{d}$, where the constant of proportionality arguably expresses how much the local fluid velocity $u_x(x)$ deviates from the substrate velocity. 
In Fig.~\ref{fig:inductiontime} we show the ratio of the (dimensionless) induction time and $(x_\mathrm{b} - x_\mathrm{d})$ calculated for our quasi one-dimensional model as a function of the substrate velocity scaled to the critical velocity $u_*$. For fast coating, the (scaled) induction time becomes constant, indicating that $u_x(x) \approx u_\mathrm{sub}$ everywhere in the immiscible region as may actually also be inferred from Fig.~\ref{fig:SimulationDomain}. Note that the solute-rich domains already redissolve before crossing the binodal, because the free energetic cost of the solute-solvent interface makes very small solute- or solvent-rich droplets energetically unfavorable. The scaled induction time levels off for high coating velocities at a value somewhat larger than unity. On the other hand, for low substrate velocities, the scaled induction time increases with decreasing velocity, indicating that the (average) fluid velocity is smaller than the substrate velocity and that this effect increases with decreasing coating velocity.

\begin{figure}[tb]
    \centering
    \includegraphics[width=0.5\textwidth]{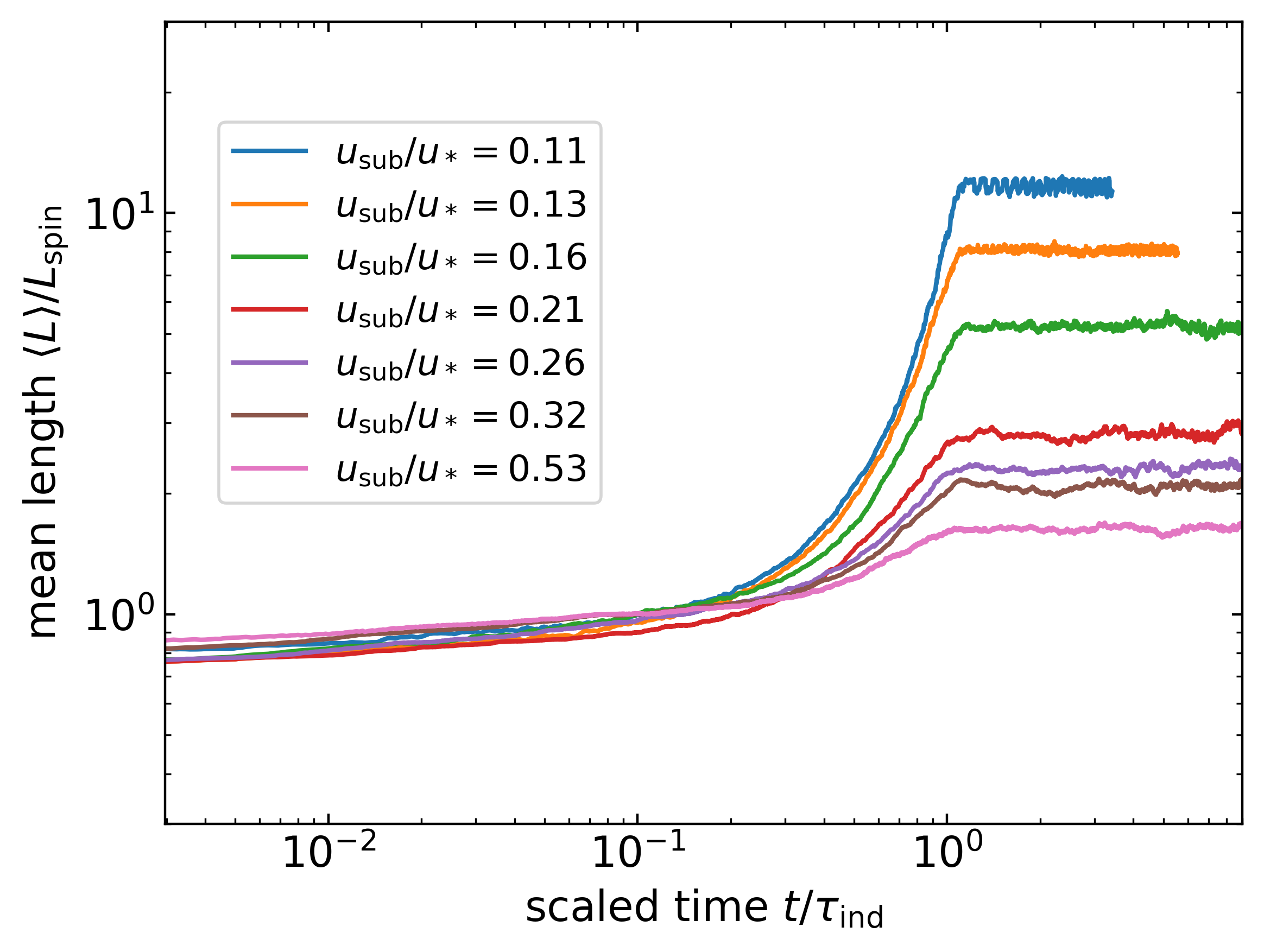}
    \caption{The mean length $\langle L \rangle$ scaled to the spinodal length $L_\mathrm{spin}$ as a function of the scaled time $t/\tau_\mathrm{ind}$ for increasing substrate velocity (downwards). The spinodal length is that for the steady-state meniscus-guided phase separation studied in the previous sections, and is equal to $2 \pi/ q_*$, with $q_*$ given by Eq.~\eqref{eq:critical_wavenumber}. The time $\tau_\mathrm{ind}$ is equal to the duration of the initial induction period. For $t/\tau_\mathrm{ind}>1$ the deposition is steady-state with mean deposition length $\langle L_\mathrm{ss}\rangle$. Conditions of the calculations correspond to the standard set in Table~\ref{tab:ParameterRanges}.}
    \label{fig:figure9B}
\end{figure}

Let us now focus on the characteristic average feature size in the immiscible region during the induction period as a function of the coating velocity. For this, we again use our quasi one-dimensional model and calculate a single mean length scale from Eq.~\eqref{eq:defmeanL}. Evidently, coarsening driven by the difference in the curvatures of the interfaces between solute-rich and poor domains, \textit{e.g.}, by Ostwald ripening, is not accounted for in these quasi one-dimensional calculations, which according to Fig.~\ref{fig:2D_early-to-late} should actually be expected to have a large influence on the mean feature size. We return to this at the end of this section. Ignoring this for the time being, we show in Fig.~\ref{fig:figure9B} the mean length $\langle L \rangle$ scaled to the spinodal length $L_\mathrm{spin}$ as a function of the time relative to the induction time. For $L_\mathrm{spin}$ we take the value for steady state coating in the $u_\mathrm{sub} \to \infty$ limit. In agreement with the discussion above, the solution remains in a mixed state for a short period of time (not shown), after which it phase separates. Again, for these short times hydrodynamic transport and evaporation do not play a role in spinodal decomposition. as we can see in Fig.~\ref{fig:figure9B}, for early times, $t/\tau_\mathrm{ind} \lessapprox 10^{-2}$, we observe that the emergent length scale is much smaller than the spinodal length $L_\mathrm{spin}$. This is to be expected because $L_\mathrm{spin}$ is the spinodal length scale for off-critical, meniscus-guided phase separation, whereas $\langle L \rangle$ is dictated by the typical feature size of the domains that form in response to the instantaneous quench throughout the immiscible region. The latter are much smaller as can in fact be seen in Figs.~\ref{fig:2D_early-to-late}A and \ref{fig:2D_early-to-late}C.

After the full immiscible region has phase separated, around about $t/\tau_\mathrm{ind} = 10^{-2}$, the influence of advection becomes important. Newly formed domains can only form at the demixing position and their characteristic feature size is set by either solute depletion or spinodal demixing, as we discussed in the Section~\ref{sec:onedim}. The mean feature size increases until $t/\tau_\mathrm{ind} = 1$, because (1) domains tend to merge, even in the quasi one-dimensional calculations due to evaporation-induced aging, (2) the new domains that form at the demixing position $x_\mathrm{d}$ are larger than those formed in response to the instantaneous quench, which remain small because (Ostwald) ripening is absent in one-dimensional calculations and (3) the domains that form in response to the instantaneous quench redissolve first. Since these domains are, on average, smaller than those that form during the steady-state meniscus-guided demixing, this results in the mean length scale to increase with time. For $t/\tau_\mathrm{ind} > 1$, we reach the dynamical steady state where the mean length becomes (approximately) constant and equal to the mean steady-state deposition length $\langle L_\mathrm{ss} \rangle$, which we also show in Fig.~\ref{fig:MeanLengthScaled} for a range of different deposition settings.

Although the induction period is governed by two different types of spinodal decomposition, we can still find universality of the time-dependent feature size $\langle L \rangle$. To show that this is the case, we argue that we may divide our spatial domain into two regions. The first region, closest to the inlet, consists of all domains that form at the demixing position $x_\mathrm{d}$ via the steady-state mechanism, and the second region consists of all domains that have formed as a response to the instantaneous quench at $t=0$. The boundary that separates the two regions moves with the fluid towards the outlet with a velocity identical to the local fluid velocity. This position is, again, also illustrated in Fig.~\ref{fig:2D_early-to-late}C with the vertical white line. Let us denote the co-ordinate associated with this boundary as $\tilde{x}(t)$, which we can calculate by inversion of $t = \int_{x_\mathrm{d}}^{\tilde{x}(t)} u_x^{-1}(x) \dint x$, and define the length of the immiscible region as $L_\mathrm{tot} = x_\mathrm{b} - x_\mathrm{d}$, with $x_\mathrm{b}$ the position of the binodal and $x_\mathrm{d}$ the demixing position. The characteristic feature size is either determined by the meniscus-guided demixing or by phase separation in response to an instantaneous quench. Hence, we can associate two separate (mean) wave numbers with the regions upstream and downstream from $\tilde{x}$. The mean wave number $\langle q \rangle(t)$ measured over the full immiscible region is the weighted average of the mean wave numbers in the separate regions, with the weight equal to the fraction of the length of the immiscible region $L_\mathrm{tot}$ corresponding to each region. Converting this mean wave number to a length $\langle L\rangle(t)$ we find for $t< \tau_\mathrm{ind}$
\begin{equation}\label{eq:UniversalCurve}
    \frac{\langle L_\mathrm{ss} \rangle}{\langle L \rangle(t)} - 1 = \Delta l(t)\left(1 - l(t)\right).
\end{equation}
where $l(t) = (\tilde{x}(t)-x_\mathrm{d})/L_\mathrm{tot}$, $\langle L_\mathrm{ss}\rangle$ is the mean length scale for steady coating that sets in for $t > \tau_\mathrm{ind}$ and depends the coating velocity, as can be seen in Figs.~\ref{fig:figure9B} and \ref{fig:MeanLengthScaled}, and $\Delta l(t) = \left(\langle L_\mathrm{ss}\rangle - L_\mathrm{ind}(t)\right)/ L_\mathrm{ind}(t)$. Here, $L_\mathrm{ind}(t)$ is the time-dependent mean length for all domains that have formed in response to the instantaneous quench, \textit{i.e.}, during the early stages of the induction period. The quantity $\Delta l(t)$ turns out to be a function of the kinetic and thermodynamic parameters that set the emergent spinodal length in the induction region.

\begin{figure}[tb]
    \centering
    \includegraphics[width=0.5\textwidth]{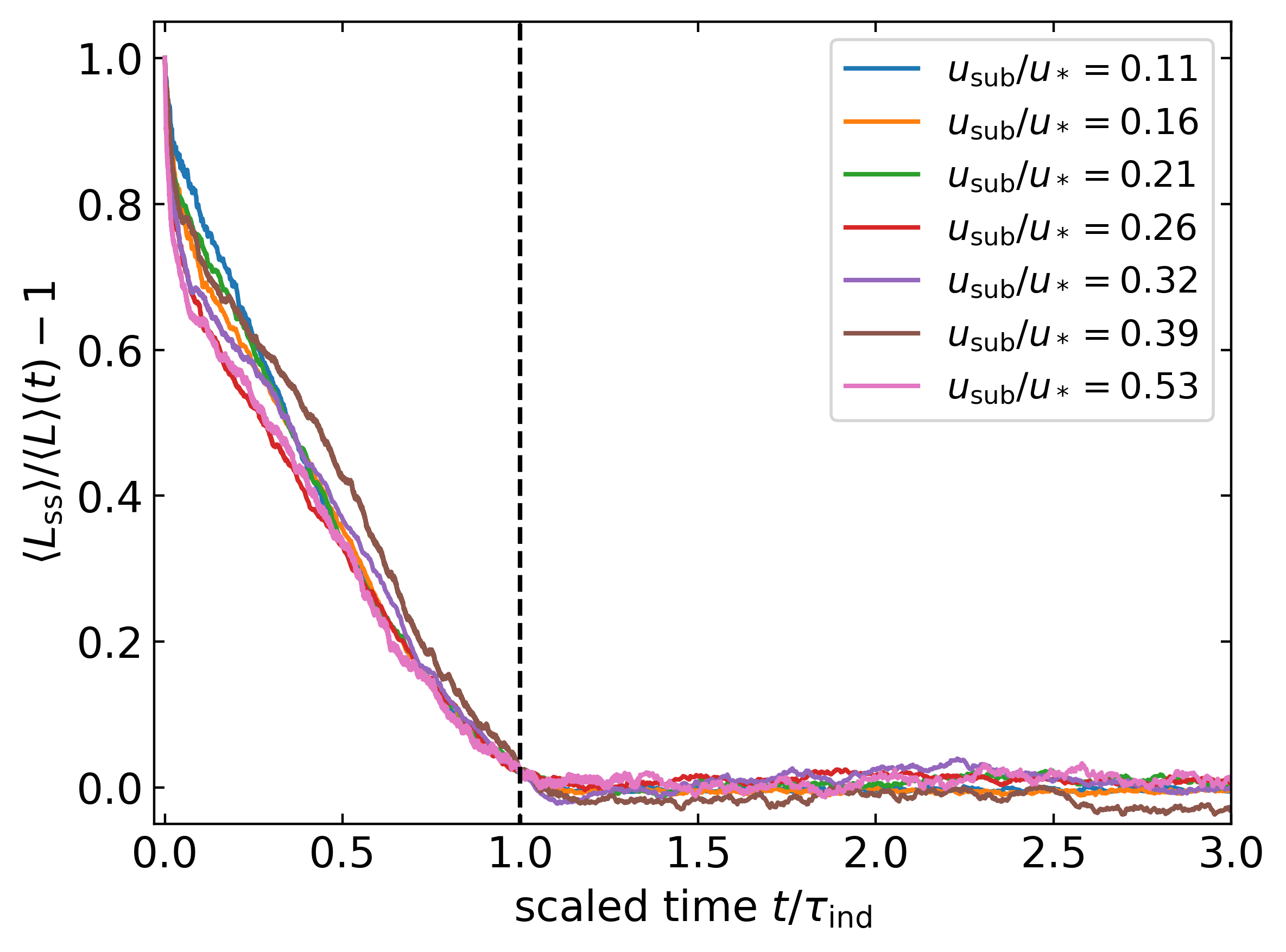}
    \caption{The steady-state length $\langle L_\mathrm{ss}\rangle$ scaled to the mean length $\langle L \rangle$ minus unity, expressed via Eq.~\eqref{eq:UniversalCurve} as a function of the time $t/\tau_\mathrm{ind}$, scaled to the duration of the induction period $\tau_\mathrm{ind}$. For $t/\tau_\mathrm{ind}\geq 1$ the deposition is in steady-state conditions. Conditions of the numerical calculations correspond to the standard set in Table~\ref{tab:ParameterRanges}.}
    \label{fig:1D_early-to-late}
\end{figure}

Making use of Eq.~\eqref{eq:UniversalCurve} allows us to more or less collapse the data onto a universal curve for all substrate velocities in the range from $u_\mathrm{sub}/u_* = 0.16$ to $u_\mathrm{sub}/u_*=0.53$, as shown in Fig.~\ref{fig:1D_early-to-late}. Evidently, while $\Delta l$ depends on time, it does so sufficiently weakly that the universal behavior is not broken for the quasi one-dimensional case. In principle, Eq.~\eqref{eq:UniversalCurve} should also hold in the quasi two-dimensional case albeit that Ostwald-type coarsening in that case becomes much more important. This expresses itself in Eq.~\eqref{eq:UniversalCurve} via the time dependence of $L_\mathrm{ind}(t)$ that is implicit in $\Delta l(t)$. Whether universality still holds for our quasi two-dimensional results remains to be seen. At present we are not able to investigate this on account of the prohibitive computation times required.

\section{Discussion and conclusion}\label{sec:discussion}
In summary, we have studied the meniscus-guided deposition of an immiscible binary solution onto a moving substrate as a model for the coating processes used in the fabrication of organic thin-film electronics. Using a numerical model centered around the lubrication approximation, we have been able to probe the relationship between the many physical parameters that control the deposition and the microscopic phase-separated dry film morphology. We find the coating velocity to be the key control parameter, and discover a morphological transition from near-isotropic phase separation to the periodic deposition of arrays of droplet-shaped solute-rich domains at right angles to the coating direction. Associated with this morphological transition, is a characteristic structural length that for sufficiently fast coating approaches the spinodal length scale and becomes independent of the coating velocity. This structural length scale increases with decreasing substrate velocity if coating is slow.

It turns out that this morphological transition can be explained in terms of the competition between diffusive and advective mass transport. Newly formed solute-rich domains accumulate solute molecules from the surrounding solution, resulting in depletion zones around them. Simultaneously, solute is supplied from the bulk fluid towards the meniscus and into the thin film at a rate that is proportional to the fluid velocity. The formation of new domains is inhibited if the growth rate of the depletion zones is higher than the supply rate of new solute molecules. This results in a longer growth time for the domains and associated depletion zones, causing the characteristic feature size to increase. In the steady state, this process perpetually repeats itself, resulting in periodic deposition of solute-rich domains. 

We quantify this process  by focusing on the two length scales fundamental to our model: the depletion length scale that characterizes the maximum size of a depletion zone and the spinodal length scale associated with phase separation in a flat and stationary reference film. The depletion length scale we show to be proportional to the ratio of the tracer diffusion coefficient and the fluid velocity, whereas the spinodal length scale introduces a fundamental minimum structure size. If the depletion length is (much) smaller than the spinodal length, phase separation is spinodal-like, and the demixing and deposition processes decouple. If the depletion length is much larger than the spinodal length, the former sets the emergent phase-separation length scale and the resulting morphology is determined by the periodic deposition. In other words, if the substrate velocity is sufficiently small, that is, on the order of the growth rate of solute-rich domains, the coating process controls the morphology of the deposited film.

The phase-separation and depletion length scales intrinsically depend on the material properties, such as the (inter)diffusion coefficient, the Flory interaction parameter, the concentration gradient stiffnesses and so on. This is not so for the underlying mechanism that gives rise to the transition in the emergent morphology, since the competition between diffusive and advective mass transport in the solvent-rich phase is \textit{always} present if a moving `miscibility front' is present~\cite{vanSaarloos2003FrontStates}. This miscibility front corresponds with the position that separates the regions in the film where the solution is (meta)stable and where it is unstable. Hence, we conclude that the morphological transition is not unique to meniscus-guided deposition and is also, perhaps surprisingly, not dependent on whether we coat in the evaporative, intermediate or Landau-Levich regime. We argue that it is a \textit{universal} transition that emerges for any coating technique where the coating is, in some sense, unidirectional, irrespective of the properties of the (binary) fluid mixture. Our results should therefore apply to other binary fluid mixtures, such as those including polymeric solutes, as well as mixtures that exhibit a rapid rise in the viscosity at high solute concentration, which could also account for the vitrification of the solute~\cite{Krieger1959ASpheres,Dreval1973ApproachSolutions,Doi1986TheDynamics}. From exploratory quasi one-dimensional calculations (not shown) we find that a concentration-dependent viscosity does indeed not affect our main conclusions. All of this is consistent with the work by Foard and Wagner~\cite{Foard2009EnslavedMixtures,Foard2012SurveyDimensions} who have surveyed with numerical calculations how a sharp or step-wise ``phase-separation front'', translating with a constant velocity, imposes the morphological evolution of a phase-separating binary fluid.

In practice, the solution rarely consists out of only two components, but typically comprises three or more, including the solvent~\cite{Franeker2015cosolvents}. The additional components can be required for the functioning of the active layer in thin-film devices. In organic photovoltaics, for example, the active layer must consist of at least an electron donor and an electron acceptor~\cite{Gaspar2018RecentCells}. Even in this case, we would expect that we can induce a morphological transition by tuning the coating velocity, as the broken symmetry in the supply of the solute material is still present. Nevertheless, the additional phase-separation pathways that are available to multi-component solutions might result in different correlations between the coating velocity and the emergent structures. This we leave for future research.

In this work, we focus attention on the initial stages of the steady-state demixing and how this couples to deposition process. While largely outside the scope of this work, the subsequent aging and coarsening of domains and how this is affected by the meniscus-guided deposition is certainly also very important for the final (dry) film morphology. From Fig.~\ref{fig:2DFinalTime} it appears that coarsening is sensitive to the coating velocity and in the high velocity case the morphology becomes isotropic much faster than in the slow velocity case. These differences can for a large part be attributed to the disparity in initial size of the domains since larger domains require much longer times to coarsen. Given the number of effects that may be at play during coarsening, we leave also a detailed study on the effects of coating on the coarsening rate for future work.

Ideally, we would directly compare our results with experimental studies. Contrary to studies concerning crystallizing solutes, experimental studies on the meniscus-guided deposition of phase-separating solutions are usually performed at coating velocities that are too high to allow for the observation of the transition in the length scales and the emergent morphology~\cite{Gu2016ComparisonCoating}. Qualitatively, however, our results should also be relevant for the early-time evolution of crystallizing solutes~\cite{Yildiz2022OptimizedCoating,Zhang2021RelationFilms,Michels2021PredictiveCoating}, even though in this case a spinodal (and hence spinodal decomposition) is believed to be absent. Namely, the mechanisms underlying the change in length scales, the moving miscibility front and the competition in mass transport are also present during the deposition of a crystallizing solute. Hence, we expect that very similar relations exist between the initial (crystallite) domain size and substrate velocity for both crystallization and liquid-liquid phase separation. Indeed, this we find to be consistent with the combined numerical and experimental findings by Michels \textit{et al.}~\cite{Michels2021PredictiveCoating} and Rogowski \textit{et al.}~\cite{Rogowski2011SolutionProperties}, where the emergent crystallite length scale increases with decreasing coating velocity. Of course, the actual morphology depends strongly on the type of transition, liquid-liquid or liquid-solid~\cite{Michels2021PredictiveCoating}.

In conclusion, we model and predict the effect of processing parameters, and in particular the coating velocity, on the thin-film morphology resulting from the meniscus-guided coating of a phase separating binary solution. We provide an intuitive explanation for the observed change in deposition morphology as a function of the coating velocity, in principle providing a concrete handle for experimental settings to control the phase separation morphology using the unidirectionality of the coating method. This transition should be universally present, independent of the chemistry of the solution, as the basic ingredients required for the observed change in the phase-separation length scale, the competition in transport processes in the solvent-rich phase and a moving miscibility front, are ubiquitously present during the unidirectional deposition of organic thin film electronics. Control over the emergent morphology can only be exerted if the coating velocity is below a certain critical velocity, below which the size of solute-rich domains increases with decreasing substrate velocity. Surprisingly, since the critical velocity is set only by the material properties and evaporation rate and not by the coating conditions, we find it to be unrelated to the evaporative or Landau-Levich regimes.

\section*{Supporting Information}
Movies of the simulation snapshots of our quasi two-dimensional calculations in S1, S2 and S3 (MP4). 

\section*{acknowledgments}
R.d.B., J.J.M. and P.v.d.S. acknowledge funding by the Institute for Complex Molecular Systems at Eindhoven University of Technology.

\bibliography{references}

\newpage
\appendix

\section*{List of Symbols}\label{app:listofsymbols}

\begin{table}[h]
    \centering
    \begin{tabular}{| c | c |}
    \hline
        Symbol & Explanation\\
        \hline
        $h_\mathrm{dry}$ & dry film height \\
        $u_\mathrm{sub}$ & substrate velocity \\
        $u_*$ & critical velocity separating evaporative and Landau-Levich regimes  \\
        $\mathbf{u}$ & height-averaged fluid velocity  \\
        $h$ & height of the solution-gas interface  \\
        $\psi$ & equivalent height of the solute  \\
        $\phi$ & volume fraction of the solute  \\
        $\eta$ & viscosity  \\
        $\kappa$ & concentration gradient stiffness  \\
        $T$ & absolute temperature\\
        $b^3$ & microscopic volume of solute\\
        $A_\mathrm{H}$ & Hamaker's constant \\
        $\sigma$ & surface tension of the solution-gas interface\\
        $k$ & evaporative mass transfer coefficient\\
        $p$ & pressure\\
        $\Delta \mu$ & exchange chemical potential (density) \\
        $f_\mathrm{loc}$ & local free energy density\\
        $\chi$ & Flory interaction parameter\\
        $\zeta$ & (Gaussian) thermal fluctuations \\
        $M$ & diffusive mobility \\
        $\omega$ & dampening factor for thermal noise\\
        $L_\mathrm{men}$ & length of static meniscus \\
        $d$ & distance between coating head and substrate \\
        $h_0$ & height at the inlet\\
        $D$ & tracer diffusivity \\
        $\tau_\mathrm{L}$ & spinodal lag time \\
        $q_*$ & emergent spinodal wave number \\
        $x_\mathrm{s}$ & position of the low-concentration spinodal \\
        $x_\mathrm{d}$ & position where the solution demixes \\
        $x_\mathrm{b}$ & position of the high-concentration binodal \\
        $\langle L \rangle$ & mean feature size \\
        $\langle L_\mathrm{SS} \rangle$ & mean feature size during steady-state demixing \\
        $L_\mathrm{ind}$ & mean feature size during induction period \\
        \hline
    \end{tabular}

\end{table}

\section{Numerics}\label{app:numerics}
We solve the lubrication equation Eq.~\eqref{eq:Lubrication:dimensionless} and the generalized diffusion equation Eq.~\eqref{eq:genDifEq:dimensionless} numerically. For reasons of convenience, we implement these equations in the so-called split form, where we solve the equations at each time step for the variables $\{h, p, \psi, \Delta \mu\}$. For the sake of numerical stability, we add additional terms to both the pressure and the free energy functional Eq.~\eqref{eq:FreeEnergy}. To the pressure $p$ we add a constant background pressure $p_0$, which does not affect the dynamics, but improves the rate of convergence. For the chemical potential density, we add a contribution to the free energy density
\begin{equation}\label{app:eq:numcontr}
    f_\mathrm{num}(\phi) = A\left(\phi^{-\gamma} + \left(1-\phi\right)^{-\gamma}\right),
\end{equation}
where $A=10^{-5}$ and $\gamma = 1$~\cite{Ronsin2022PhaseFieldFilms,Negi2018SimulatingInvestigation,Schaefer2016StructuringEvaporation}. This contribution to the free energy is relatively standard in the phase field community~\cite{Ronsin2022FormationSimulations,Schaefer2016StructuringEvaporation,Negi2018SimulatingInvestigation}. It penalizes volume fractions (very) close to pure conditions ($\phi = 0$ or $\phi = 1$) allowing for larger time steps, while not affecting the calculations qualitatively.

As discussed in the main text, we adopt a multi-step approach to efficiently simulate the meniscus-guided deposition of an immiscible solution. First, we simulate the (one-dimensional) deposition of a miscible solution on a coarse grid of $N = 5000$ grid points with grid spacing $\Delta x = 4 \times 10^{-1}$ \textmu m until the (dynamic) steady-state solutions to Eqs.~\eqref{eq:Lubrication:dimensionless}~and~\eqref{eq:genDifEq:dimensionless} are found. The one-dimensional steady-state profiles are shown in Fig.~\ref{fig:SimulationDomain} in blue (solution layer height) and red (volume fraction). 

Next, we interpolate the coarse grid to a finer grid by refining it by a factor 100, such that the new grid spacing $\Delta x$ allows for at least 5 grid points within phase boundaries, which is required for stable calculations~\cite{Wodo2011ComputationallyProblem}. For the two-dimensional calculations, we extend this grid into the $y$-direction. In the finer grid simulation, we change the Flory interaction parameter $\chi$ and the stiffness $\kappa$ to the (range of) values presented in Table~\ref{tab:ParameterRanges}, effectively replacing the initial ideal solution by a non-ideal solution. We determine the positions corresponding with the volume fractions for the low solute concentration spinodal (left-most vertical dashed line in Fig.~\ref{fig:SimulationDomain}) and the high solute concentration binodal (right-most vertical dashed line in Fig.~\ref{fig:SimulationDomain}). Only between these two positions do we expect that the phase-separated domains are present. The numerical domain extend at least $5000$ grid points beyond the spinodal position to prevent the boundary conditions to affect our results. Adding more grid points does not affect our results. We extract the upstream boundary conditions from the coarse grid of the ideal solution, and set the downstream boundary conditions to Neumann conditions. Boundary conditions are implemented using ghost nodes.

We discretize all gradients using second-order central finite differences and integrate time with the semi-implicit Euler method. The deterministic part of the Eqs~\eqref{eq:Lubrication:dimensionless}~and~\eqref{eq:genDifEq:dimensionless} we integrate implicitly, the stochastic contributions due to the thermal noise ($\zeta$ in Eq.~\eqref{eq:genDifEq:dimensionless}) explicitly using the method by Schaefer \textit{et al.}~\cite{Schaefer2016StructuringEvaporation}. Our model is implemented in parallel using the PETSc library~\cite{Abhyankar2018PETSc/TS:Library,Balay2024PETScPage,Balay2024PETSc/TAOManual} and the thermal fluctuations are calculated using the pseudo-random number generator ``Xoshiro256+''~\cite{Blackman2021ScrambledGenerators}. We use adaptive time steps, using the number of Newton-Raphson cycles required for convergence as control parameter: for 10 or fewer iterations the timestep is increased by 20\%, otherwise the time step is decreased by 20\%. If more than 25 Newton-Raphson cycles are required, we discard the time step and reattempt after halving the time step. This adaptive time step method is efficient for calculations of spinodal decomposition~\cite{Wodo2011ComputationallyProblem}.

For the one-dimensional calculations we use a domain decomposition method based on the additive Schwarz method with a single overlap node, and use the ILU(1) preconditioner on each block, similar to numerical method of Ref.~\cite{Zheng2015A3D}. For the two-dimensional calculations we use the PETSc Fieldsplit preconditioner. The generalized diffusion equation Eq.~\eqref{eq:genDifEq:dimensionless} is solved again using a additive Schwarz method with a single overlap node with ILU(2) preconditioner for each block. The lubrication equation Eq.~\eqref{eq:Lubrication:dimensionless} is solved with the direct sparse LU solver MUMPS~\cite{Amestoy2001AScheduling}.

\section{Early time analysis}\label{app:earlytime}
The early time behavior of spinodal decomposition is related to the linear stability of Eqs.~\eqref{eq:genDifEq:dimensionless}~and~\eqref{eq:Lubrication:dimensionless}. The results presented in this Appendix follows the work by Schaefer \textit{et al.}~\cite{Schaefer2016StructuringEvaporation}, but differ because we must linearize around a (steady-state) reference state as $h(\mathbf{r}) = h_0(\mathbf{r})$, $\phi(\mathbf{r},t) = \phi_0(\mathbf{r}) + \delta \phi(\mathbf{r},t)$ and $\psi(\mathbf{r},t) = \psi_0(\mathbf{r}) + \delta \psi(\mathbf{r},t)$ with $\delta\phi \ll  \phi_0$ and $\delta\psi \ll \psi_0$. We presume that spatial variations in the reference state are small over the length scale relevant for spinodal decomposition, $\nabla h_0(\mathbf{r}),\nabla \phi_0(\mathbf{r}) \ll 1$ and that variations in the velocity $\mathbf{u}$ are small also, such that we can neglect the gradients of these quantities. In essence, this is similar to a small gradient approximation underlying the lubrication approximation.

The resulting linearized diffusion advection equation reads
\begin{align}
&\dpart{\delta\psi}{t} + \nabla\cdot\left(\mathbf{u} \delta\psi\right)= \\
&\nabla \cdot\left(M(x)\nabla\left[\dparth{f_\mathrm{loc}}{\phi}{2}\nabla \delta \psi - \kappa \nabla^3 \delta\psi\right]\right). \nonumber
\end{align}
Next, we switch to a co-moving reference frame, by following a single volume element. Practically, to do so, we must subtract $\mathbf{u} \cdot \nabla \delta\psi$ on the left hand side. This is obviously only exact if the velocity field is constant, but we deem it a reasonable approximation for a weakly-varying velocity field. 

In such a co-moving reference frame within a `local' approximation, all quantities that depend on position via the height or solute height now only depend on time (which can be translated to spatial position). Hence, we find
\begin{equation}
\dpart{\delta\psi}{t} = \nabla \cdot\left(M(t)\left[\dparth{f_\mathrm{loc}}{\phi}{2}(t) \nabla \delta\psi - \kappa \nabla^3 \delta\psi\right]\right).
\end{equation}
Switching to Fourier modes, we finally obtain
\begin{equation}
\ddif{\delta\psi}{t} = \left(M(t)\left[\dparth{f_\mathrm{loc}}{\phi}{2}(t) q^2 \delta\psi + q^4\kappa \delta\psi\right]\right).
\end{equation}
The solution of this ordinary differential equation is trivial, and reads
\begin{equation}\label{eq:appB:solution}
\ln\left(\frac{\widehat{\delta\psi}(q,t)}{\widehat{\delta\psi}(q,0)}\right) = -R(q,t) t.
\end{equation}
with
\begin{equation}
R(q,t) = -q^2 \frac{1}{t}\int_{0}^t \dint t'\left(M(t') \left[\dparth{f_\mathrm{loc}}{\phi}{2}(t') + q^2 \kappa \right]\right)
\end{equation}

We find the dominant wave number by setting the first derivative of the previous equation to zero,
\begin{equation}\label{eq:app:criticalwavenumber}
q_*^2(T) = -\frac{\int_{0}^t \dint t' M(t') \dparth{f_\mathrm{loc}}{\phi}{2}(t')}{2 \kappa \int_{0}^t \dint t'M(t')},
\end{equation}
and the lag time by inserting $q = q_*(\tau_\mathrm{L})$ and $t = \tau_\mathrm{L}$ in Eq.~\eqref{eq:appB:solution}, which yields 
\begin{equation}
r = \frac{\kappa}{2} q_*^4 \int_{0}^{\tau_\mathrm{L}} \dint t'M(t') =  \frac{1}{8 \kappa} q_*^4 \frac{\left(\int_{0}^{\tau_\mathrm{L}} \dint t' M(t') \dparth{f_\mathrm{loc}}{\phi}{2}(t')\right)^2}{\int_{0}^{\tau_\mathrm{L}} \dint t' M(t')}.
\end{equation}
Here, $r = \ln\left(\widehat{\delta\psi}(q_*,\tau_\mathrm{L})/\widehat{\delta\psi}(q_*,0)\right)$ a quantity that sets the relative magnitude for the amplitude fluctuations we associate with the lag time.

Since demixing occurs close to the position of the spinodal, we introduce the linearization for the local free energy around the position of the low-concentration branch of the spinodal
\begin{equation}\label{eq:B8}
    \dparth{f_\mathrm{loc}}{\phi}{2} \approx f_{\phi\phi\phi} \alpha t',
\end{equation}
where $f_{\phi\phi\phi} = \dparth{f_\mathrm{loc}}{\phi}{3}$ evaluated at the position of the spinodal, $\alpha = \dpart{\phi}{t}$, \textit{i.e.}, the rate of change of the volume fraction \textit{due to solvent evaporation}, evaluated at the position of the low concentration branch of the spinodal. Note that the zeroth order contribution $\dparth{f_\mathrm{loc}}{\phi}{2}\big|_\mathrm{spin} = 0$. Inserting Eq.~\eqref{eq:B8} in the definition of the lag or amplification time, we find
\begin{equation}\label{eq:app:tauL2}
r = \frac{\left(f_{\phi\phi\phi}\alpha\right)^2}{8\kappa} \frac{\left(\int_{0}^{\tau_\mathrm{L}} \dint t'M(t') t'\right)^2}{\left( \int_{0}^{\tau_\mathrm{L}} \dint t' M(t')\right)}.
\end{equation}

Using the definition of the mobility Eq.~\eqref{eq:doubledeg}, replacing the volume fractions by $\phi = \phi_\mathrm{spin} + \alpha t'$, and linearizing $\tau_L$, we find that to lowest order, 
\begin{equation}
r = \frac{(f_{\phi\phi\phi}\alpha)^2 M(\phi_\mathrm{spin})}{32\kappa } \tau_L^3,
\end{equation}
or after reshuffling some term and introducing $x_\mathrm{d} \approx u_x \tau_\mathrm{L}$, we find
\begin{equation}
x_L^3 =  \left(r\frac{32 \kappa}{(f_{\phi\phi\phi})^2 M(\phi_\mathrm{spin})}\right)\left(\langle u_x \rangle\big|_\mathrm{spin.}\right)^3\alpha^{-2},
\end{equation}
where only the final two terms might vary between different coating conditions. Hence, we observe that either the effective evaporation rate via $\alpha$ or the coating velocity influence the position where domains emerge. 

Using this expression for the amplification time, we can straightforwardly evaluate Eq.~\eqref{eq:app:criticalwavenumber}, yielding to lowest order in $\tau_\mathrm{L}$
\begin{equation}
    q_* \propto \left(\left|f_{\phi\phi\phi} \alpha\right| \frac{\tau_\mathrm{L}}{4 \kappa}\right)^{1/2} \propto \alpha^{1/6}.
\end{equation}

\end{document}